\documentclass[fleqn,usenatbib]{mnras}

\usepackage{newtxtext,newtxmath}
\usepackage{float}
\usepackage{xcolor}

\usepackage[T1]{fontenc}

\DeclareRobustCommand{\VAN}[3]{#2}
\let\VANthebibliography\thebibliography
\def\thebibliography{\DeclareRobustCommand{\VAN}[3]{##3}\VANthebibliography}

\usepackage{graphicx}	
\usepackage{amsmath}	
\usepackage{siunitx}    
\usepackage{physics}    
\usepackage{caption}
\usepackage{orcidlink}

\newcommand{\pc}{\>{\rm pc}}
\newcommand{\kpc}{\mbox{$\>{\rm kpc}$}} 
\newcommand{\kmsk}{\mbox{$\>{\rm km\, s^{-1}\, kpc^{-1}}$}}
\newcommand{\Gyr}{\mbox{$\>{\rm Gyr}$}}
\newcommand{\Myr}{\mbox{$\>{\rm Myr}$}}

\newcommand{\rin}{R_\mathrm{in}}
\newcommand{\rout}{R_\mathrm{out}}

\usepackage{xspace}
\newcommand{\Gaia}{\textit{Gaia}\xspace}

\title[\textit{Deep Potential} in barred galaxies]{Recovering the gravitational potential in a rotating frame: \textit{Deep Potential} applied to a simulated barred galaxy}

\author[T. Kalda et al.]{
Taavet Kalda\,\orcidlink{0000-0002-0417-8645},$^{1}$\thanks{E-mail: taavet.kalda@gmail.com}
Gregory M. Green\,\orcidlink{0000-0001-5417-2260},$^{1}$
Soumavo Ghosh\,\orcidlink{0000-0002-6549-7455}$^{1}$
\\
$^{1}$ Max-Planck-Institut f\"{u}r Astronomie, K\"{o}nigstuhl 17, D-69117 Heidelberg, Germany
}

\date{Accepted XXX. Received YYY; in original form ZZZ}

\pubyear{2015}

\begin{document}
\label{firstpage}
\pagerange{\pageref{firstpage}--\pageref{lastpage}}
\maketitle

\begin{abstract}
  Stellar kinematics provide a window into the gravitational field, and therefore into the distribution of all mass, including dark matter. \textit{Deep Potential} is a method for determining the gravitational potential from a snapshot of stellar positions in phase space, using mathematical tools borrowed from deep learning to model the distribution function and solve the Collisionless Boltzmann Equation. In this work, we extend the \textit{Deep Potential} method to rotating systems, and then demonstrate that it can accurately recover the gravitational potential, density distribution and pattern speed of a simulated barred disc galaxy, using only a frozen snapshot of the stellar velocities. We demonstrate that we are able to recover the bar pattern speed to within \SI{15}{\%} in our simulated galaxy using stars in a \SI{4}{kpc} sub-volume centered on a Solar-like position, and to within \SI{20}{\%} in a \SI{2}{kpc} sub-volume. In addition, by subtracting the mock ``observed'' stellar density from the recovered total density, we are able to infer the radial profile of the dark matter density in our simulated galaxy. This extension of \textit{Deep Potential} is an important step in allowing its application to the Milky Way, which has rotating features, such as a central bar and spiral arms, and may moreover provide a new method of determining the pattern speed of the Milky Way bar.
\end{abstract}




\section{Introduction}

One of the major goals of Milky Way dynamics is the recovery of the gravitational potential. This is because the gravitational potential is sourced by all forms of matter, both baryonic and dark. As far as is known at present, the dark component can only be mapped by its gravitational effects (direct or indirect). Recovering the gravitational potential is thus important for mapping the dark matter distribution in the Milky Way. Motivated by the ongoing \Gaia mission, which is providing six-dimensional phase-space information for tens of millions of stars \citep{Gaia16, GaiaDR323}, we tackle this problem using a data-driven method using mathematical tools from deep learning, which we term ``\textit{Deep Potential}'' \citep{Green20,Green23}.

The dynamics of the trajectories of the stars in the Milky Way are dominated by gravitational forces. While we are able to measure the positions and velocities of the stars, stellar accelerations due to the Milky Way potential -- typically on the order of \SI{1}{cm.s^{-1}.yr^{-1}} -- are exceedingly difficult to directly measure with present-day instruments \citep{silverwood_easther_2019}. Some systems do lend themselves to acceleration measurements, but they are either in the presence of strong accelerations or exotic systems such as pulsars or eclipsing binaries \citep{Ghez00, Chakrabarti21, Phillips21}.

This means we are effectively limited to a single snapshot of the positions and velocities of the stars in the Milky Way. We describe this snapshot via its distribution function, $f(\vec{x}, \vec{v})$, which gives the density of stars in phase space (position and velocity). However, unless additional assumptions are made about the nature of a gravitational system, any gravitational potential is consistent with any snapshot of the distribution function, as the gravitational potential only determines the evolution of the system. To connect the distribution function to the potential, one typically assumes the system to be in a steady state, in which the distribution function does not vary with time. In this work, for the first time, we weaken this assumption and require only that stationarity hold in some arbitrarily rotating frame. We achieve this by concurrently finding both the potential and the rotation parameters which render the distribution function most stationary. This is important because many physical systems of interest, such as the Milky Way, contain rotating features. The Milky Way harbors a central bar \citep[e.g.,][]{LisztandBurton1980,Binneyetal1991,Weinberg1992,Binneyetal1997,BlitzandSpergel1991,Hammersleyetal2000,WegandGerhard2013,Shen20} and spiral arms \citep[e.g.,][]{Oort1958,GeorgelinandGeorgelin1976,Gerhard2002,Churchweletal2009,Reidetal2014}, both of which break the standard, non-rotating stationarity assumption. The properties of the Milky Way bar are still not fully determined, with the two leading models being a slowly rotating long bar or a fast-rotating short bar \citep[e.g.,][]{Clarke22}. The method we present in this paper should be sensitive to the pattern speed of the Milky Way bar or local spiral features, and could help determine whether the bar has a slow or fast pattern speed.

Existing methods for modeling the dynamics typically rely on taking velocity moments of the Collisionless Boltzmann Equation via Jeans modeling \citep{BinneyTremaine2008} or rely on simplified models for the distribution function \citep[e.g.,][]{Schwarzschild79, Syer96, McMillan08, Bovy13, Magorrian14}. This has been historically motivated by either the lack of or small quantities of available full six-dimensional phase-space data, but also by Milky Way by and large being axisymmetric or having other symmetries which lends itself to aforementioned modeling approaches. However, with the large quantitative and qualitative improvements brought by \Gaia, capturing the full complexity of the stellar kinematic data -- including non-axisymmetric structures -- has become more feasible. \textit{Deep Potential} goes beyond simple parametric models and borrows several techniques from the realm of deep learning. With the methodological improvements developed in this work, \textit{Deep Potential} makes the following minimal assumptions about the underlying physics of the kinematic system:
\begin{enumerate}
    \item The motions of stars are guided by a gravitational potential $\Phi(\vec x)$.
    \item We observe the phase-space coordinates of stars (the ``kinematic tracers'') that are statistically stationary in a rotating frame.
    \item The overall matter density is non-negative everywhere: $\rho(\vec x) \geq0$. We can express this via gravitational potential using the Poisson equation as $\rho(\vec x) = \laplacian \Phi/(4\pi G) \geq 0$.
\end{enumerate}
Notably, we do not assume that the gravitational potential is sourced by the observed kinematic tracers, as other matter components (\textit{e.g.}, dark matter) can contribute to the potential.

Our previous work on this method outlined its theoretical motivations and demonstrated its effectiveness on simpler toy models with observational errors and non-stationarities \citep{Green20,Green23}. These toy models involved drawing particles from analytic models of stationary systems or evolving a set of tracer particles in a fixed background potential. In this work, we move further and demonstrate the method on a self-consistent $N$-body simulation of a disc galaxy with a prominent central bar, and we further relax the assumption that the distribution function be stationary in the observed, ``laboratory'' frame of reference. Furthermore, recent work by \citet{Ghosh+Roadmap2023} demonstrated that the presence of a prominent bar produces systematic biases in recovering the underlying distribution function and gravitational potential when using action-based dynamical modeling.

Similar methods to \textit{Deep Potential}, using normalizing flows to represent the stellar distribution function and then determining the gravitational potential by assuming stationarity, have been developed by \citet{An2021}, \citet{Naik2022GalacticAcc}, and \citet{Buckley2022GalacticDM,Buckley23}. \citet{Lim2023} recently applied normalizing-flow-based modeling to a sample of Milky Way stars in order to estimate the local dark matter density. The major qualitative addition made by our present work over previous methods is the relaxation of the stationarity assumption, to hold in an arbitrarily rotating frame. This is an important advance in order to accurately model the Milky Way, which harbors rotating features such as a central bar.

In this paper, we describe and expand on the methodology of \textit{Deep Potential} (Section~\ref{sec:method}), outline the $N$-body simulation and the selection of the dataset (Section~\ref{sec:bar_model}), test the performance of the method (Sections~\ref{sec:df} and \ref{sec:phi}) and discuss future prospects (Section~\ref{sec:conclusions}).

\section{Method}
\label{sec:method}
This work builds on the ``\textit{Deep Potential}'' method, which is explained in \citet{Green20} and \citet{Green23}. In this work, we extend \textit{Deep Potential} to allow for concurrent fitting of both the gravitational potential and the rotating frame in which the system appears most stationary. In a barred galaxy, for example, this could correspond to a frame rotating at the pattern speed of the bar. Here, we briefly review the key components of the \textit{Deep Potential} method and then derive a generalized stationarity assumption in a rotating frame.

The first assumption of \textit{Deep Potential} is that stars orbit in a background gravitational potential, $\Phi (\vec x)$. The density of an ensemble of stars in six-dimensional phase space (position $\vec x$ and velocity $\vec v$) is referred to as the distribution function, $f(\vec x, \vec v)$. The evolution of the distribution function is described by the Collisionless Boltzmann Equation:
\begin{equation}
    \dv{f}{t} = \pdv{f}{t} + \sum_{i} \left(v_i\pdv{f}{x_i} - \pdv{\Phi}{x_i}\pdv{f}{v_i}\right) = 0.
\end{equation}

Our second assumption is that the distribution is stationary. In previous work, we assumed that stationarity holds in the laboratory frame (\textit{i.e.}, the frame in which the positions and velocities of the stars are measured, such as the barycenter of the Solar System): $\pdv*{f}{t} = 0$. However, disc galaxies typically have rotating features, such as bars and spiral arms. In a barred galaxy, for example, it would be reasonable to assume that the galaxy would appear more stationary in a frame that co-rotates with the bar, instead of in an inertial frame. We therefore generalize the stationarity condition to a frame that is rotating with with angular speed $\vec \Omega$ around an axis passing through a point $\vec x_0$ in space. We additionally allow the stationary frame to be move with constant velocity $\vec v_0$ relative to the laboratory frame. In the Milky Way, $\vec x_0$ and $\vec v_0$ could represent the location and velocity of the Galactic Center relative to the Solar System, and $\vec \Omega$ (directed along the rotation axis of the Galaxy) could represent the pattern speed of either the central bar or of the spiral arms. Here, we work out the stationarity condition for the general case when $\vec x_0 \neq$, $\vec v_0 \neq \vec 0$ (for real observations, the location of the rotation axis and the velocity of the center of system are not always zero). In general, the parameters describing the stationary frame can either be fixed, or can be determined concurrently with the gravitational potential. Though we derive the stationarity condition in full generality, in the numerical experiments in this work, we will later fix $\vec{x}_0$ and $\vec{v}_0$ to zero.

In the following, we denote the partial derivative of the distribution function w.r.t. time \textit{in the rotating frame} as $(\pdv*{f}{t})_\Omega$.
The generalized stationarity condition states that this partial derivative should equal zero. By translating the rotating-frame partial derivative to partial derivatives in the laboratory frame, we obtain our generalized stationarity condition in terms of laboratory-frame quantities:
\begin{align}
    \label{eqn:generalized-stationarity}
    \left(\pdv{f}{t}\right)_\Omega &=
      \pdv{f}{t}
      + \sum_{i}\left( u_i(\vec x) \pdv{f}{x_i}
      + w_i(\vec v)\pdv{f}{v_i}\right)
    = 0,\\
    \vec u(\vec x) &= \vec\Omega\times (\vec x - \vec x_0) + \vec v_0,\\
    \vec w(\vec v) &= \vec\Omega\times(\vec v - \vec v_0).
\end{align}
For a full derivation of this transformation, see Appendix~\ref{app:stationarity}.
Combining this with the Collisionless Boltzmann Equation, we arrive at
\begin{align}
    \label{eq:CBE+stat}
    \left(\pdv{f}{t}\right)_\Omega
    = \sum_i \left[
      (u_i - v_i)\pdv{f}{x_i}
      + \left(\pdv{\Phi}{x_i} + w_i\right)\pdv{f}{v_i}
    \right]
    = 0.
\end{align}
If the distribution function is truly stationary, the gravitational potential can be uniquely determined by solving Eq.~\eqref{eq:CBE+stat}. Realistic physical systems will, however, not be completely stationary, and as such, there may not exist any potential which would render the system stationary (See \citealt{An2021} and \citealt{Green23} for discussion). In general, therefore, \textit{Deep Potential} recovers the potential which minimizes some measure (to be discussed below) of the total non-stationarity in the system. Note that we do not assume that the gravitational potential is sourced by the observed stellar population alone. Accordingly, we do not impose the condition
\begin{equation}
    \laplacian\Phi(\vec x) = 4\pi G \int f(\vec x, \vec v)\dd[3]{\vec v}.
\end{equation}

\begin{figure*}
    \includegraphics[width=\linewidth]{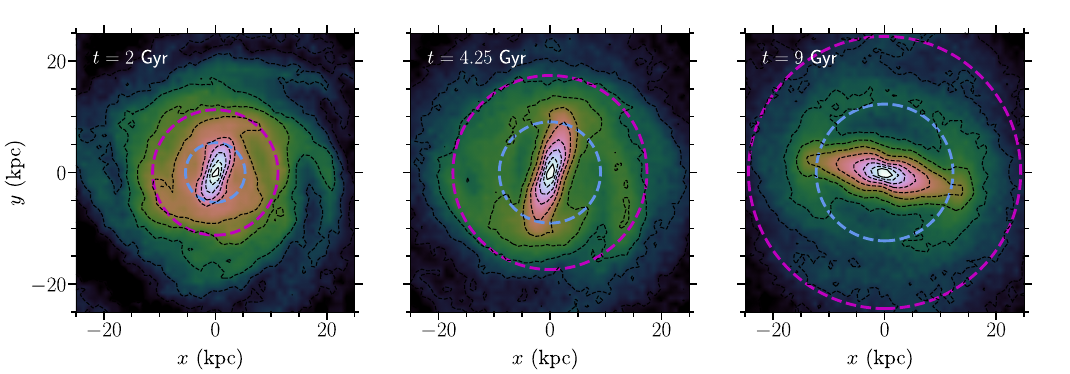}
    \caption{Face-on distribution of all stellar particles, calculated at $t = 2$, $4.25$, and $9 \Gyr$ for our fiducial bar model. The dashed black curves denote the constant density contours. The blue circle in each panel denotes the extent of the bar ($R_{\rm bar}$) whereas the magenta circle denotes the location of the corotation radius ($R_{\rm CR}$). The galaxy harbors a prominent bar, which grows in radial extent from the earliest to the latest time step.}
    \label{fig:density_maps}
\end{figure*}

\subsection{Modeling the distribution function}

In practice, when we observe stellar populations, we obtain a discrete sample of points in phase space, rather than a smooth distribution function $f(\vec x,\vec v)$. In order to obtain gradients of the underlying distribution function, we require a continuous, differentiable object representing $f(\vec x, \vec v)$. For this purpose, we use normalizing flows, which are a class of algorithms used for density estimation in unsupervised machine learning (for a review, see \citealt{Kobyzev2019}). A normalizing flow works by learning a set of invertible coordinate transformations that turn a simple distribution, usually a normal distribution, into a more complex one that fits the observed data. The complexity of the distributions it can capture is limited by the number of parameters describing the coordinate transformations. There are many approaches for constructing normalizing flows, and the field is in constant development. In this work, we opt to use FFJORD \citep{Grathwohl18}, though the particular choice of normalizing flow method is not critical to the working of \textit{Deep Potential}. The main drawback of normalizing flows is that most implementations assume the training data to be continuous everywhere. This, however, is not a significant problem, as most stellar systems exhibit the same property.

Given a sample of $n$ stars with positions $\vec x_i$ and velocities $\vec v_i$, we train a normalizing flow $f(\vec x, \vec v)$ using stochastic gradient descent to obtain the parameters of the flow that maximize the log-likelihood
\begin{align}
    L_f = \sum_{i=1}^{n} \ln f\left(\vec x_i, \vec v_i\right). 
\end{align}
The loss is supplemented with Jacobian and kinetic regularization \citep{Finlay2020}, as detailed in Section~\ref{sec:implementation}. When doing subsequent analysis, we also multiply the output of the normalizing flow by $n$. This is because normalizing flows are normalized to one, but we're interesting in the output being the number density of training data. The great advantage of using a normalizing flow is that our representation is both highly flexible and auto-differentiable. When implemented in a standard deep-learning framework, such as TensorFlow \citep{tensorflow2015-whitepaper}, PyTorch \citep{NEURIPS2019_9015} or JAX \citep{jax2018github}, it is possible to automatically differentiate the distribution function at arbitrary points in phase space, in order to obtain the terms $\pdv*{f}{\vec x}$ and $\pdv*{f}{\vec v}$ in the Collisionless Boltzmann Equation.

\subsection{Modeling the gravitational potential}
After learning the distribution function, we find the gravitational potential $\Phi(\vec x)$ and angular rotation speed $\Omega$ that best satisfy the Collisionless Boltzmann Equation and generalized stationarity assumption given in Eq.~\eqref{eq:CBE+stat}. To parameterize the gravitational potential, we use a feed-forward neural network which takes as input a three-dimensional vector $\vec x$ and outputs a scalar, $\Phi$.
We concurrently train the parameters of the potential and $\Omega$ to minimize
\begin{equation}
    \label{eq:grav_loss_int}
    L_\Phi = \int
      \mathcal L \left[
        \left(\pdv{f(\vec x,\vec v)}{t}\right)_\Omega
        \!\!\! , \,
        \laplacian \Phi(\vec x)
      \right]
      f(\vec x, \vec v)
      \dd[3]{\vec x}\dd[3]{\vec v},
\end{equation}
where $\mathcal L$ is the differential contribution to the loss of an individual point in phase space, given by
\begin{equation}
    \label{eq:grav_loss_individual}
    \mathcal L =
      \sinh^{-1}\left[
        \alpha\left|\left(\pdv{f}{t}\right)_\Omega\right|
      \right]
    + \lambda \sinh^{-1}\left(
        \beta \max\left\{-\laplacian\Phi, 0\right\}
      \right).
\end{equation}
The first term penalizes non-stationarity in a frame rotating with angular speed $\Omega$ while the second term penalizes negative mass densities. We first take the absolute value of $(\pdv*{f}{t})_\Omega$, in order to penalize positive and negative changes in the phase-space density equally. The inverse hyperbolic sine function down-weights large values, while the constant $\alpha$ sets the level of non-stationarity at which our penalty transitions from being approximately linear to being approximately logarithmic. The loss is supplemented with $\ell_2$ regularization based on the neural-network weights that describe $\Phi(\vec x)$. The integral in Eq.~\eqref{eq:grav_loss_int} is computationally expensive to evaluate directly, but can be approximated by averaging $\mathcal{L}$ over $m$ samples drawn from the distribution function, where $m$ is a sufficiently large number:
\begin{equation}
    \label{eq:grav_loss_sum}
    L_\Phi \approx
      \frac{n}{m} \sum_{i=1}^{m}
        \mathcal{L} \left[
          \left(\pdv{f(\vec x_i,\vec v_i)}{t}\right)_\Omega
          \!\!\! , \,
          \laplacian \Phi(\vec x_i)
        \right].
\end{equation}
The constant $n/m$ comes from the normalization of the distribution function, and can be omitted when implementing the loss function.

\subsection{Implementation}
\label{sec:implementation}
In this paper, we implement \textit{Deep Potential} in TensorFlow~2 \citep{tensorflow2015-whitepaper} and using TensorFlow Distributions \citep{Dillon17}. All of our code is publicly available under a permissive license that allows reuse and modification with attribution, both in archived form at \url{https://doi.org/10.5281/zenodo.8390759} 
and in active development at \url{https://github.com/gregreen/deep-potential}.

To represent the distribution function, we use a chain of three FFJORD normalizing flows, each with eight or twenty densely connected hidden layers, depending on the system, of \texttt{hidden\_size} neurons (in this paper, we use $\mathtt{hidden\_size} = 128$ or 256 for different systems) and a $\tanh$ activation function. For our base distribution, we use a multivariate Gaussian distribution with mean and variance along each dimension set to match the training data set. During training, we impose Jacobian and kinetic regularization with strength \num{5e-4} \citep{Finlay2020}, which penalizes overly complex flow models and tends to reduce training time. We train our flows using the rectified Adam optimizer \citep{Liu2019}, with a batch size of $2^{13}$ (8096). We find that this relatively large batch size leads to faster convergence (in wall time) than more typical, smaller batch sizes. We begin the training with a ``warm-up'' phase that lasts 2048 steps, in which the learning rate linearly increases from 0 to \num{0.001}. Thereafter, we use a constant learning rate. We decrease the learning rate by a factor of two whenever the training loss fails to decrease \num{0.01} below its previous minimum for 512 consecutive steps (this period is termed the “patience”). We terminate the training when the loss drops below \num{e-6}.

After training our normalizing flow, we draw $m = 2^{21}$ ($\sim 2$ million) phase-space coordinates, and calculate the gradients $\pdv*{f}{\vec x}$ and $\pdv*{f}{\vec v}$ at each point (using auto-differentiation), for use in learning the gravitational potential.

We represent the gravitational potential using a feed-forward neural network with four densely connected hidden layers, each with 512 neurons and a tanh activation function (we eschew more commonly used activation functions, such as ReLU, which have discontinuous derivatives, as these may lead to unphysical potentials). The network takes a three-dimensional input (the position $\vec x$ in space), and produces a scalar output (the potential). No activation function is applied to the final scalar output. We add in an $\ell_2$ loss on the potential network weights with strength $0.1/\mathtt{n\_weights}$, where \texttt{n\_weights} is the total number of weights in the network. We train the network using the rectified Adam optimizer, with batches of $2^{15}$ (\num{32768}) phase-space coordinates. We use a similar learning-rate scheme as before, with a warm-up phase lasting 2048 steps, an initial learning rate of \num{0.001}, and a patience of 2048 steps. In the potential loss function (Eq.~\ref{eq:grav_loss_individual}), we set $\alpha=\num{1e5}$, $\beta=1$, and $\lambda=1$, which affect the penalties on non-stationarity and negative gravitational mass densities.

When fitting both the distribution function and the gravitational potential, we reserve \SI{25}{\%} of our input data as a validation set. After each training step, we calculate the loss on a batch of validation data, in order to identify possible overfitting to the training data. Such overfitting would manifest itself as a significantly lower training loss than validation loss. In the experiments in this paper, no significant overfitting is observed—the difference in the likelihoods of the training and validation sets is typically less than 1\%.

The choices made here are by no means set in stone, and can
be altered without changing the overall \textit{Deep Potential}
framework. In particular, rapid advances have been made in
research into normalizing flows over the preceding years. As more accurate and/or computationally efficient flows are
developed, they can be used by \textit{Deep Potential}.


\section{Fiducial $N$-body bar model}
\label{sec:bar_model}

\begin{figure}
  \centering
  \includegraphics[width=\linewidth]{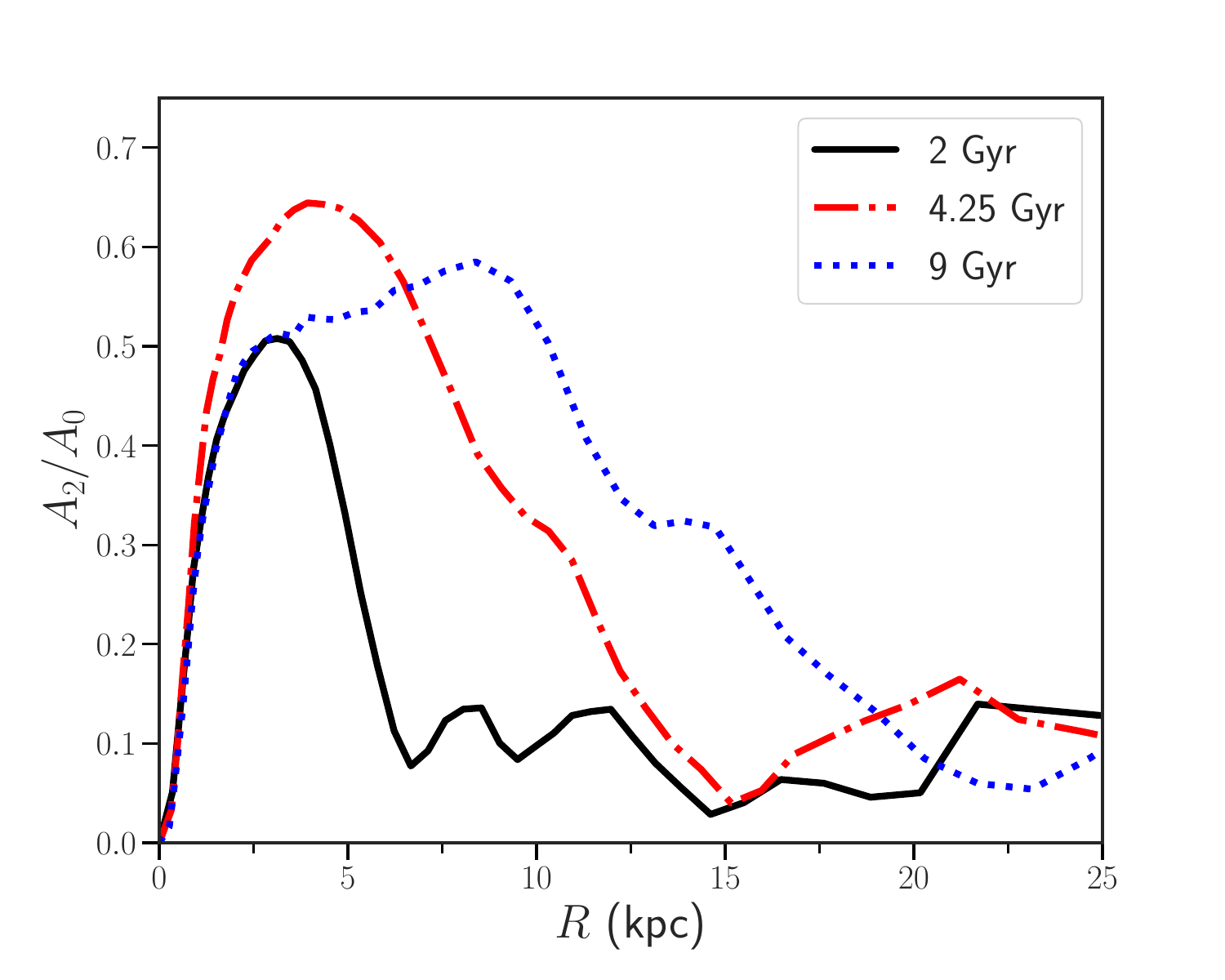}
  \caption{A measure of the radial extent of the bar in our simulated galaxy at three time steps. At each time step, we measure the radial profile of the $m=2$ Fourier coefficient of stellar density in cylindrical annuli (normalized by the $m=0$ coefficient) using Eq.~\ref{eq:fourier_calc}. A prominent peak in the radial profile of the $m=2$ Fourier coefficient clearly demonstrates the presence of a central bar at all three time steps. The central bar extends radially over time, becoming increasingly prominent in the stellar mass distribution.}
  \label{fig:m2Fourier_profiles}
\end{figure}

\begin{table}
  \centering
  \caption{Key structural parameters for the initial, equilibrium model of our simulated galaxy. This model is subsequently integrated in a self-consistent manner for \SI{9}{Gyr}, during which time it develops a central bar. See Section~\ref{sec:bar_model} for details of the $N$-body simulation.}
  \begin{tabular}{ccccccc}
    \hline
    $R_{\rm d}$ & $h_{\rm z}$ & $M_{\rm disc}$ & $R_{\rm H}$ & $M_{\rm dm}$ & $n_{\rm disc}$ & $n_{\rm dm}$\\
    (kpc) & (kpc) & ($\times 10^{11} M_{\odot}$) & (kpc) & ($\times 10^{11} M_{\odot}$) & ($\times 10^{5}$) & ($\times 10^{5}$)\\
    \hline
    4.7 & 0.3 & 1 & 10 & 1.6 & 10 & 5\\
    \hline
  \end{tabular}
  \label{table:key_param}
\end{table}

To demonstrate \textit{Deep Potential} in a rotating frame, we make use of an $N$-body simulation of a collisionless stellar disc that subsequently develops a strong bar in the central region. In Section~\ref{sec:sim_setup}, we describe the initial equilibrium set-up, and the basic structural parameters pertaining to the fiducial bar model. In Section~\ref{sec:bar_properties}, we explain some of the key bar properties which will be used later in this work. Finally, in Section~\ref{sec:dataset_selection}, we specify the tracers that we use for training our normalizing flows. 

\subsection{Simulation set-up and equilibrium configuration}
\label{sec:sim_setup}

The details of the initial equilibrium model were previously explained in \citet{Ghoshetal2022} (See \texttt{rthick0.0} there). Here, for the sake of completeness, we briefly mention the equilibrium set-up for our fiducial bar model.

The equilibrium model consists of an axisymmetric stellar disc which is embedded in a live dark matter halo. The stellar disc is modeled with a Miyamoto-Nagai profile \citep{MiyamatoandNagai1975} with a potential of the form
\begin{equation}
  \Phi_{\rm disc} =
    -\frac{
      GM_{\rm disc}
    }{
      \sqrt{R^2+\left(R_{\rm d}+\sqrt{z^2+h_{\rm z}^2}\right)^2}
    }
    \, ,
  \label{eq:pot_disc}
\end{equation}
where $R_{\rm d}$ and $h_{\rm z}$ are the characteristic disc scale length and scale height, respectively, and $M_{\rm disc}$ is the total mass of the stellar disc. The dark matter halo is modeled with a Plummer sphere \citep{Plummer1911}, with potential of the form
\begin{equation}
  \Phi_{\rm dm} (r) = -\frac{GM_{\rm dm}}{\sqrt{r^2+R^2_{\rm H}}} \, ,
  \label{eq:pot_dm}
\end{equation}
where $R_{\rm H}$ is the characteristic scale length, and $M_{\rm dm}$ is the total halo mass. Here, $r$ and $R$ are the radius in the spherical and the cylindrical coordinates, respectively. In Table~\ref{table:key_param}, we list the key values of the structural parameters for stellar disc as well as the dark matter halo. The total number of particles used to model each of these structural components is also listed in Table~\ref{table:key_param}.

The initial conditions of the stellar disc are obtained using an iterative method \citep[See][]{Rodionovetal2009}. For this model, we constrain the density profile of the stellar disc while allowing the velocity dispersions (both vertical and radial components) to adjust such that the system converges to an equilibrium solution \citep[for details, see][]{Fragkoudietal2017,Ghoshetal2022}. The simulation is run using a TreeSPH code by \citet{SemelinandCombes2002}. A hierarchical tree method \citep{BarnesandHut1986} with opening angle $\theta = 0.7$ is used for calculating the gravitational force, which includes terms up to quadrupole order in the multipole expansion. In addition, we use a Plummer potential for softening the gravitational forces with a softening length $\epsilon = 150 \pc$. The equations of motion are integrated using a leapfrog algorithm \citep{Pressetal1986} with a fixed time step of $\Delta t = 0.25 \Myr$. The model is evolved for a total time of $9 \Gyr$.

\subsection{Properties of the stellar bar}
\label{sec:bar_properties}

To study the robustness of the \textit{Deep Potential} in a rotating frame when applied to a barred galaxy, we choose three snapshots, namely at $t = 2 \Gyr$, $4.25 \Gyr$, and $9 \Gyr$ from the fiducial bar model. Fig.~\ref{fig:density_maps} shows the corresponding face-on density distribution of stellar particles at these three time steps. A visual inspection reveals that at all three time steps the model harbors a prominent stellar bar in the central region. 

We quantify the strength of the bar by taking the Fourier decomposition (in azimuthal angle $\phi$) of the density in annuli (of cylindrical radius $R$). We then calculate the strength of the normalized $m=2$ Fourier coefficient as a function of $R$:
\begin{equation}
  \left. \frac{A_m}{A_0} \right|_R = \left|
    \frac{
      \sum_j M_j e^{\mathrm{i}m\phi_j}
    }{
      \sum_j M_j
    }
  \right| \, ,
  \label{eq:fourier_calc}
\end{equation}
where $A_m$ denotes $m$-th coefficient of the Fourier moment of the stellar density distribution, $M_j$ is the mass of the $j^{\mathrm{th}}$ particle (not to be confused with the Fourier coefficient), and $\phi_j$ is the corresponding azimuthal angle. For each radius $R$, the summation runs over all particles within the radial annulus $[R, R+\Delta R]$, with $\Delta R = 0.5 \kpc$. Fig.~\ref{fig:m2Fourier_profiles} shows the corresponding radial profiles of the $m=2$ Fourier coefficient at the three selected time steps. We see that at $t = 2 \Gyr$, the bar is rapidly forming while at $t= 4.25 \Gyr$ the bar reaches its maximum strength (i.e., highest peak value of the $m=2$ Fourier coefficient). By the end of the simulation run at $t=9 \Gyr$, the bar remains strong. For a detailed exposition of the temporal growth of the bar, the reader is referred to \citet{Ghoshetal2022}.

\begin{table}
  \begin{tabular}{l|llll}
    \hline
    shell name & $R_\mathrm{in}$ & $R_\mathrm{out}$ & $n$ & $n_\mathrm{with\_padding}$    \\\hline
    1          & \SI{0}{kpc}     & \SI{2}{kpc}    & 123997 & 168647 \\
    2          & \SI{2}{kpc}   & \SI{5}{kpc}   & 124433 & 228354 \\
    3          & \SI{5}{kpc}   & \SI{10}{kpc}   & 146839 & 315165 \\
    4          & \SI{10}{kpc}  & \SI{16}{kpc}   & 129019 & 327434 \\\hline
    combined   & \SI{0}{kpc}     & \SI{16}{kpc}   & 524288 &  \\\hline
  \end{tabular}
  \caption{At each time step, the simulated galaxy is divided into spherical shells, which are separately modeled using normalizing flows and then stitched together. Above, we describe the shells used for partitioning the training volume at time step $t=\SI{2}{Gyr}$. The radii $R_\mathrm{in}$ and $R_\mathrm{out}$ of the inner and outer cylindrical surfaces of each of the shells is given, along with the number of tracer particles $n$ inside each of them. $n_\mathrm{with\_padding}$ refers to the number of particles after padding the shells with additional ``virtual particles.'' These virtual particles are added in order to transform the sharp inner and outer boundaries of the shells into a smooth roll-off. This eases the normalizing flow training.}
  \label{table:shells}
\end{table}

The remaining two properties of the bar that are relevant to this work are the extent of the bar, $R_{\rm bar}$, and the pattern speed of the bar, $\Omega_{\rm bar}$. Following \citet{GhoshDiMatteo2023}, at time $t$, we define $R_{\rm bar}$ as the location where $A_2/A_0$ drops to 70\%\ of its peak value. The corresponding extent of $R_{\rm bar}$ is indicated in Fig.~\ref{fig:density_maps} by a blue circle. We measure the bar pattern speed by fitting a straight line to the temporal variation of the phase-angle of the $m=2$ Fourier mode. This method assumes that the bar rotates rigidly with a single pattern speed in that time-interval. The bar slows with time, with bar pattern speeds of $20.7 \kmsk$, $12.15 \kmsk$, and $8.1 \kmsk$ at 2, 4.25 and 9~Gyr, respectively. A rotating bar induces a number of resonances, namely, corotation (CR), the Inner Lindblad Resonance (ILR) and the Outer Lindblad Resonance (OLR). To determine the locations of these resonaces, we first need to compute the radial variation of the circular velocity (equivalently, the rotation curve). At time $t$, the circular velocity $v_{\rm c}$ is calculated with the asymmetric drift correction \citep{BiineyTremaine2008}:
\begin{equation}
  v_{\rm c}^2 =
    v_{\phi}^2 + \sigma_{\phi}^2
    - \sigma_{R}^2 \left(
      1 + \dv{\ln\rho}{\ln R} + \dv{\ln\sigma^2_R}{\ln R}
    \right)
  \, .
  \label{eq:asy_drift1}
\end{equation}
Here, $v_{\phi}$ is the azimuthal velocity, whereas $\sigma_{R}$ and $\sigma_{\phi}$ denote the radial and the azimuthal velocity dispersion, respectively. Using the rotation curve, we calculate the radial location of the CR, defined by $\Omega (R = R_{\rm CR}) = \Omega_{\rm bar}$. The corresponding $R_{\rm CR}$ values are indicated in Fig.~\ref{fig:density_maps}. We also provide the numerical values of $R_\mathrm{bar}$ and $R_\mathrm{CR}$ in Table~\ref{table:omegas}. As the bar in our fiducial model slows down with time, the location of the CR is pushed farther out in the disc.

\begin{figure*}
    \centering
    \includegraphics[width=0.9\linewidth]{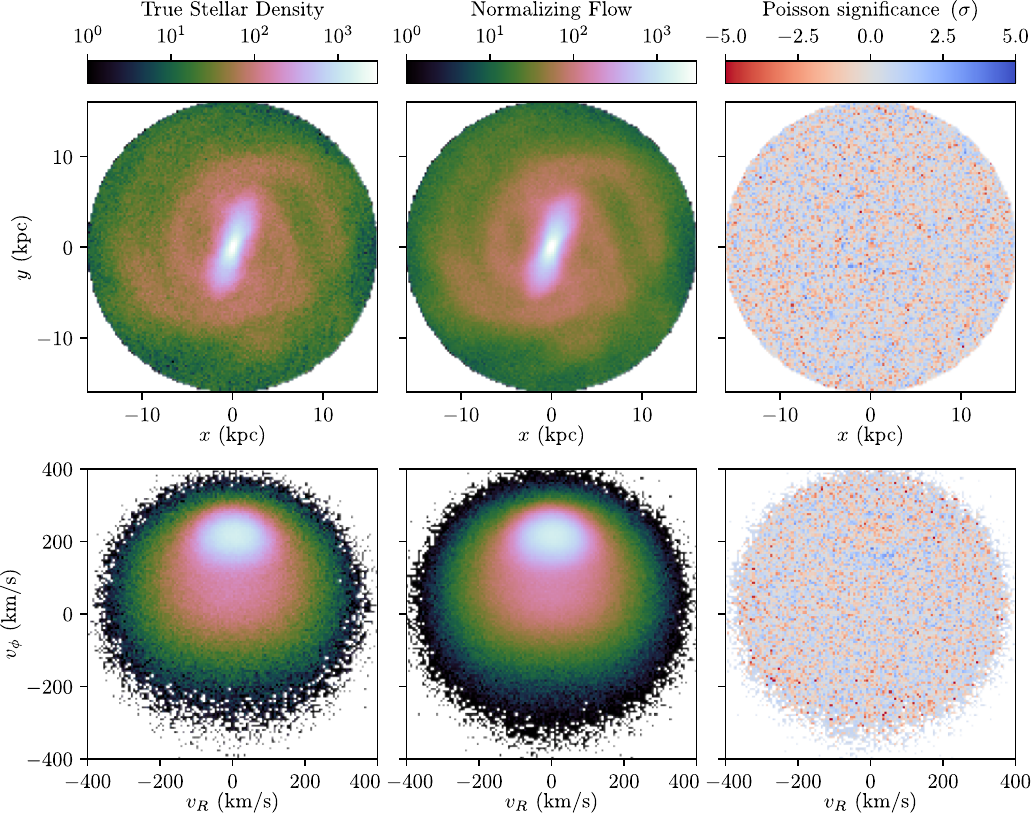}
    \caption{A demonstration of the performance of our normalizing flow model of the stellar phase-space distribution function in two projections. At time $t=\SI{2}{Gyr}$, we plot 2D histograms of selected stellar particles (left column) and of $2^{21}=2097152$ synthesized particles sampled from the trained normalizing flow (middle column), and a comparison between the two (right column). The top row shows a face-on view of the galaxy, while the lower panel shows one velocity-space projection ($v_{\phi}$ vs. $v_R$) of the galaxy. The density of the synthesized particles has been renormalized by an overall constant to match the density of the stellar particles. For each bin, we define the Poisson significance as being $(n_\mathrm{NF} - n_\mathrm{data})/\sqrt{n_\mathrm{data}}$, where $n_\mathrm{NF}$ is the renormalized number of samples in the bin drawn from the normalizing flow and $n_\mathrm{data}$ is the number of stellar particles in the same bin. As can be seen above, our normalizing flow captures all prominent features in the galactic stellar distribution, including the central bar and spiral arms. We thus obtain a smooth, differentiable representation of the galactic stellar population.}
    \label{fig:df_t80}
\end{figure*}

\subsection{Dataset selection}
\label{sec:dataset_selection}

In order to create samples of kinematic tracers for \textit{Deep Potential}, we randomly select $n=2^{19}$ (\num{524288}) stellar particles at each time step, from inside a cylindrical volume of radius $R=\SI{16}{kpc}$ (centered at the origin of the system) and half-height (in $z$) of $H=\SI{2}{kpc}$.

At each time step, we orient the coordinate system such that the $z$-axis is parallel to the angular momentum of the stellar particles and the origin corresponds to the peak density of the bar. Choosing the peak density as the origin is important, because there is a displacement in the order of \SI{100}{pc} between the peak density and the center of mass of the stellar particles of the system. This is most likely caused by the system being in a transient state and having differential rotation between the central and outer regions.

Each stellar particle has a mass of $\SI{92000}{M_\odot}$. To represent the composite particles as a collection of individual stars, we would need to upsample each composite particle by assuming that the density follows a Plummer sphere with $\varepsilon=\SI{150}{pc}$ (this is caused by the softening length). This can however introduce artificial clumping that can manifest itself as spurious densities or accelerations inferred from the gravitational potential. With this in mind, we instead treat the stellar particles as individual stars, and do not perform any upsampling. 

We also note that as the method currently stands, the training data in the chosen volume is assumed to have uniform completeness (\textit{i.e.}, each star has the same probability of being selected). If the completeness were not uniformly complete, the spatially varying selection function would introduce spurious gradients into the distribution function and hence invalidate the stationarity assumption. Uniform completeness is easy to guarantee for in the mock dataset, but with real observations, it is influenced by multiple factors such as crowding, dust, color and scanning patterns. Hence, for real Milky Way datasets, we expect modeling of the selection function to be critical.

\section{Normalizing flow}
\label{sec:df}

\begin{figure}
    \centering
    \includegraphics[width=0.9\linewidth]{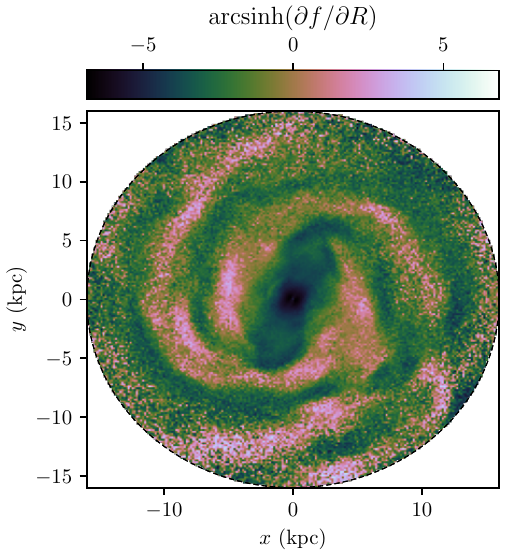}
    \caption{A 2D projection of the radial gradients in stellar density at time $t=\SI{2}{Gyr}$. We draw a sample of $2^{21}$ particles, and take the median $\pdv*{f}{R}$ in $x-y$ bins. We compress the color scale by applying the $\mathrm{arcsinh}$ function. As evidenced by the smoothness of the radial gradients, our stitching process, in which we fit the distribution function of cylindrical annuli separately and the join them together, does not introduce prominent boundary effects (\textit{i.e.}, discontinuities).}
    \captionsetup{singlelinecheck=false, justification=justified}
    \label{fig:DF_df_dR_x_y_t80}
\end{figure}

\begin{figure*}
    \centering
    \includegraphics[page=1, width=0.9\linewidth]{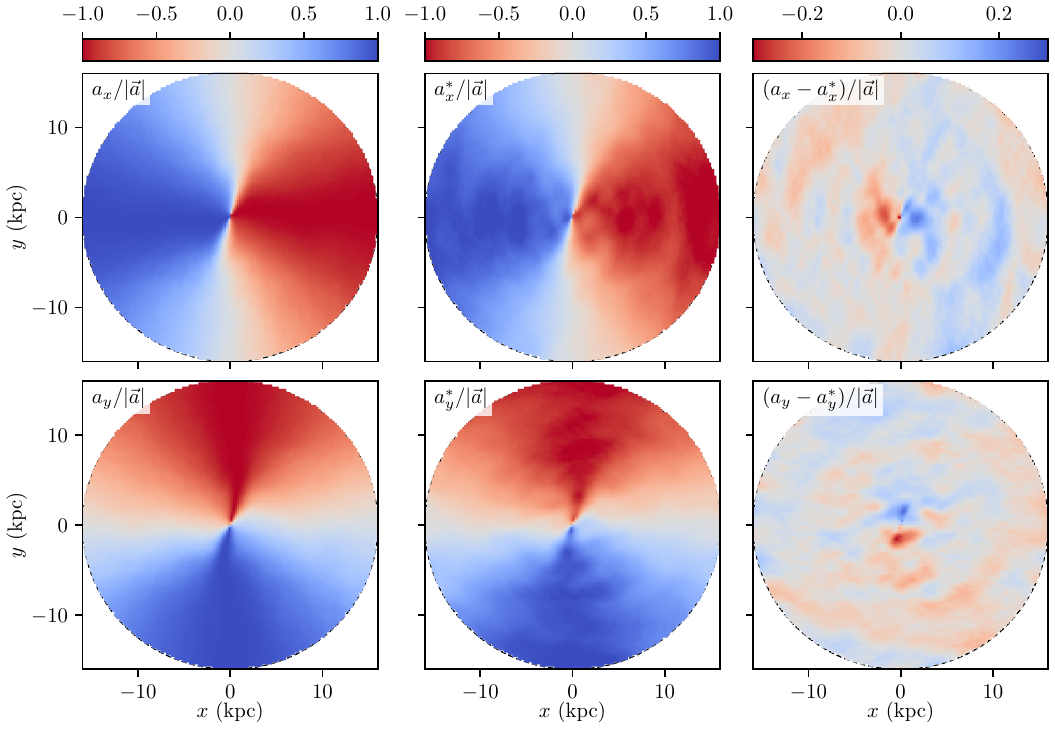}
    \caption{A comparison of the gravitational accelerations inferred by \textit{Deep Potential} in the midplane of our simulated galaxy with the true accelerations. For the snapshot at $t=\SI{2}{Gyr}$, we plot ground-truth accelerations in the $z=0$ plane obtained from the simulation (left column), recovered accelerations inferred from the gravitational model (middle column), and a comparison between the two (right column). At each point in the plane, we normalize the accelerations by the modulus of the true acceleration vector, so that the plotted values are always of order unity. We only plot the $x$ and $y$ components of acceleration, as the $z$-component at $z=0$ is very close to zero, and out of the plane is similar to the density estimates from the gravitational potential. In the midplane, we recover the overall smooth pattern of accelerations over the disc, with small-scale fluctuations at the level of five percent.}
    \label{fig:forces_t80}
\end{figure*}

\subsection{Constructing the normalizing flow}
\label{sec:constructing_df}

When training the flows for the different time steps, three limitations of our FFJORD implementation must be addressed. The first challenge arises from the fact that the densities of the tracer particles vary by up to three orders of magnitude between the central and outer regions of the training volume. These large variations in density cause systematic biases in volumes of high density. To address this, we split the training volume into four concentric cylindrical shells, such that the variation in the density of the tracer particles in each shell is lessened. This means training a separate normalizing flow for each shell. Each concentric shell is bounded by an inner surface with cylindrical radius $R=R_\mathrm{in}$, an outer surface with cylindrical radius $R=R_\mathrm{out}$, and two flat surfaces at $z=\pm H = \pm \SI{2}{kpc}$. For details on the number of particles in each shell for the first time step, see Table~\ref{table:shells} (the other two time steps are treated in a similar fashion).

The second limitation comes from the difficulty that FFJORD has in capturing discontinuities in the training data. When crossing the boundary of the training volume, the density of the tracer particles discontinuously drops from a finite value (inside the volume) to zero (outside the volume), causing the first and higher derivatives of the distribution function to be undefined. FFJORD outputs a continuous normalizing flow that attempts to model the discontinuity, but ends up introducing systematic biases in the boundary region. The third limitation comes from the known difficulty of FFJORD in capturing volumes that are not topologically equivalent to spheres (in this case, a cylindrical shell is topologically equivalent to a donut). These latter two limitations are addressed by adding additional ``virtual'' particles outside the boundary, such that the volume inside $R<R_\mathrm{in}, |z|<H$ is filled with a roughly uniform density of particles and the distribution function drops smoothly to zero outside the cylinder defined by radius $R_\mathrm{out}$ and half-height $H$. We train a normalizing flow with the virtual particles included. Subsequently, to draw a sample from one of the shells, we reject the sample that fall outside the chosen volume. For more technical details about the virtual particles, see Appendix~\ref{app:df_padding}.

In the end, we retrieve a sample of $m=2^{21}=\num{2097152}$ points from the desired distribution function by drawing a proportionate number of samples from each of the four cylindrical shells separately and concatenating them together. The resulting sample is then used for training the potential. 

\subsection{Validation}

The most straightforward diagnostic of the trained normalizing flows is a comparison of their predicted phase space density with the training data. In Fig.~\ref{fig:df_t80}, we compare two-dimensional projections of our normalizing flow and the true stellar distribution function for one time step ($t = \SI{2}{Gyr}$). As can be seen, there are no visible systematic biases in the recovered phase space density, even in regions of low and high density. We observe similar behavior for the other two time steps.

Since the N-body simulation consists of discrete particles, rather than a smooth phase-space density (even though an individual particle is spatially a Plummer sphere, it is a delta function in velocity space), it is not possible to directly compare the gradients $\pdv*{f}{\vec x}$, $\pdv*{f}{\vec v}$ of the trained flow with its corresponding true values. However, the stitching procedure provides a surprising diagnostic that proves to be useful. At the boundaries between two neighboring shells, both the densities and the gradients of the flows should match, as their combination must represent a smooth distribution function. We can use this as a test of how well the normalizing flow performs at the boundaries. This effect is clearest when plotting a face-on projection of $\pdv*{f}{R}$ in the mock galaxy, as in Fig.~\ref{fig:DF_df_dR_x_y_t80}. We note that this diagnostic serves to supplement the two-dimensional projections of the phase space density, not replace it. Even when the phase space densities do not show obvious systematics, we find signs of discontinuities in $\pdv*{f}{R}$. There are also cases in which the reverse is true. With our final normalizing flow implementation and stitching procedure, such discontinuities are not readily apparent.


\section{Gravitational Potential}
\label{sec:phi}
After obtaining a total of $m=2^{21}$ samples for each time step, including the gradients $\pdv*{f}{\vec x}$ and $\pdv*{f}{\vec v}$, we train a potential neural network $\Phi(\vec x)$ and rotation speed $\Omega$ (for each time step) that best minimizes the aforementioned loss function in Eq.~\ref{eq:grav_loss_sum}. In general, we find training the potential to be more robust and straightforward than training the normalizing flows. In particular, given a sample of distribution function gradients, different choices for the structure and hyperparameters of the potential neural network yield very similar results.

Given the gravitational potential, it is possible to compare the predicted accelerations and densities (from the gradients and Laplacian of the potential, respectively) with the ground truth obtained from the simulation itself. Importantly, $\Phi(\vec x)$ represents the contribution from all forms of matter, including dark matter, so the comparisons with the simulation must be performed accordingly.

\subsection{Modeled rotation speed}
The rotation speeds that \textit{Deep Potential} captures at different time steps are listed in Table~\ref{table:omegas}. These can be directly compared with the rotation speed of the bar (as measured from the $m=2$ Fourier mode of stellar density in the galactic midplane; see Section~\ref{sec:bar_properties}), as the bar is the most significant rotating feature present in the system. Despite the large non-stationarities and secular evolution of the system, we capture the rotation speed to within $\sim$\SI{3}{km.s^{-1}.kpc^{-1}}, or $\sim$\SI{20}{\%} at all time steps. \textit{Deep Potential} thus recovers the bar rotation speed using only a frozen snapshot of the stellar kinematics.

\begin{table}
  \begin{tabular}{l|l|l|l|l}
    \hline
    $t$ & $R_{\rm bar}$ & $R_{\rm CR}$ & $\Omega_\mathrm{bar}$ & $\Omega$ \\
    (Gyr) & (\si{kpc}) & (\si{kpc}) & (\si{km.s^{-1}.kpc^{-1}}) & (\si{km.s^{-1}.kpc^{-1}})\\
    \hline
    2 & 5 & 10.5 & 20.7 & 17.5 \\
    4.25 & 8 & 17.5 & 12.15 & 14.2 \\
    9 & 11.5 & 23.8 & 8.1 & 10.4
  \end{tabular}
  \caption{For all three time steps of our simulated galaxy, we compare the rotation speed of the bar ($\Omega_{\rm bar}$, calculated by considering the time-evolution of the $m=2$ Fourier mode of stellar density, as outlined in Section~\ref{sec:bar_properties}) with the best-fit value for the rotation speed $\Omega$ of the frame shift which renders the system most stationary, according to \textit{Deep Potential}. We also provide the values for the radius ($R_{\rm bar}$) and corotation radius ($R_{\rm CR}$) of the bar. In all three time steps, we recover the pattern speed of the bar to within $\sim$\SI{3}{km.s^{-1}.kpc^{-1}}, corresponding to a maximum fractional error of $\sim$\SI{20}{\%}.}
  \label{table:omegas}
\end{table}

\subsection{Comparison of the modeled accelerations}

We obtain the accelerations predicted by the gravitational model by calculating the gradients of the model via auto-differentiation: $\vec a = -\vec \nabla \Phi(\vec x)$. We can compare this with the ground truth from the simulation by summing over the contributions of the particles in the $N$-body simulation, both stellar and dark matter (and accounting for the softening length). Fig.~\ref{fig:forces_t80} compares the prediction with the ground truth at time step $t = \SI{2}{Gyr}$. In most of the galaxy, we recover accelerations to within $\sim$\SI{20}{\%}. The largest discrepancies occur near the center of the galaxy, where the bar is strongest. However, the major axis of the bar is almost exactly reproduced, and can be immediately identified in the spatial pattern of $a_x$ and $a_y$.

\begin{figure*}
    \centering
    \includegraphics[width=\linewidth]{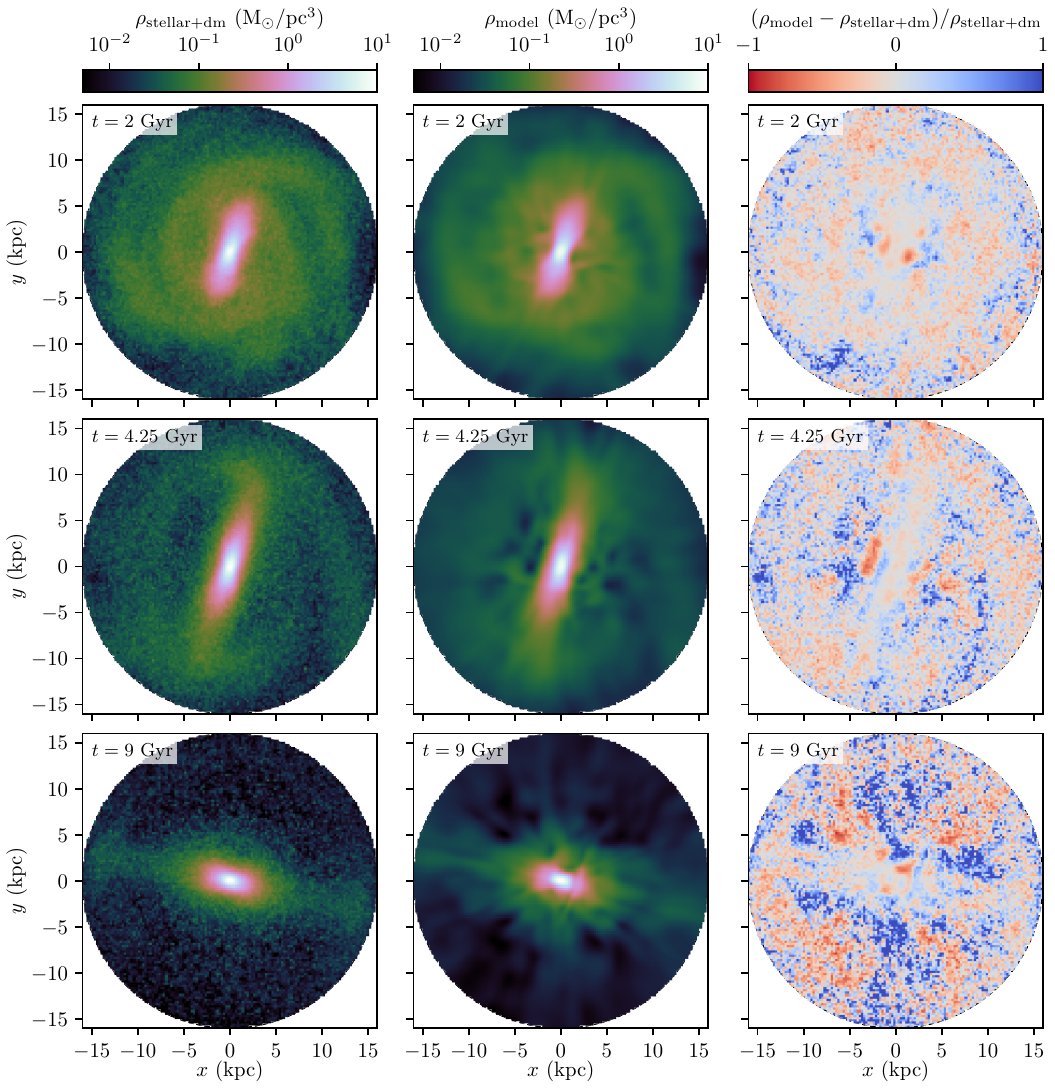}
    \caption{Comparison of the true total mass density in the plane of our simulated galaxy with the density recovered by \textit{Deep Potential}. In the left column, we plot the total density at the three time steps ($t = 2$, 4.25, and $\SI{9}{Gyr}$), calculated by summing the densities from the stellar and dark matter particles. The middle column shows the density implied by the gravitational potential that \textit{Deep Potential} recovers from the stellar kinematics. In the right column, we plot the fractional difference between the recovered and the true density. At each time step, we recover the major features of the galaxy, including the central bar and spiral arms. Though the true density varies over several orders of magnitude over the plotted region of the galaxy, we generally recover the density to a few tens of percent, with the residuals taking the form of spatially oscillating (with scale $\sim$\SI{1}{kpc}) fluctuations about zero. We note that the density is most accurately recovered when the bar is weakest (at $t = 2$ and $\SI{4.25}{Gyr}$), and that the largest density residuals tend to occur at small galactic radii, along the bar minor axis.}
    \label{fig:rho_x_y}
\end{figure*}

\subsection{Comparison of the modeled density}
\label{sec:densities}

We obtain model densities by taking the Laplacian of the modeled gravitational potential, once again by auto-differentiaton: $\rho_\mathrm{model} = \laplacian\Phi/(4\pi G)$. Because $\Phi(\vec x)$ represents the overall gravitational potential, sourced by both stellar and dark matter particles, the density estimate also represents the total density. We can compare this with the ground truth from the simulation, by summing over the contributions of all the particles. Due to the softening length of the simulation, each particle represents a Plummer sphere density with scale length $\varepsilon = \SI{150}{pc}$.

The model vs. ground truth comparison for all time steps is shown in Fig.~\ref{fig:rho_x_y}. We see that the major features of the simulated galaxy are reproduced, including the central bar and spiral arms. In general, the model performs worse in the presence of low densities (See the outer regions at $t=\SI{9}{Gyr}$) and high density gradients, such as when moving along the semi-minor axis of the bar near the galactic center. We also observe ubiquitous small-scale fluctuations in the predicted
density. Identifying the origin of these fluctuations is difficult, as the effects of the potential causes are difficult to disentangle. One possible cause could be the inherent non-stationarity of the data, stemming from the system being in a state of evolution. This effect could be compounded by the smoothing length of each individual particle manifesting itself as spurious densities over a characteristic scale that is comparable to the smoothing length ($\varepsilon = \SI{150}{pc}$). Finally, discrepancies in the accurate representation of the underlying distribution function by the normalizing flow could also contribute. Particularly, in the last two time steps, the hyperparameters for the normalizing flow are not tuned as thoroughly as for the first time step.

The effects from non-stationarity can be lessened by smoothing the density maps with some suitable kernel, as done in \citet{Buckley23}, for example. This is useful when we are interested in the average density in a neighborhood of some particular point. However, we do not focus on smoothed densities in this work, as it does not provide significant additional insight into the performance of our rotating-frame framework.

By considering the modeled density and subtracting the ground-truth baryonic density, we can build an estimate for the dark matter density in the system. This approach is motivated by the fact that stellar density can be estimated from direct observations, while dark matter is not directly observable. For example, in systems such as the Milky Way, there are analytic models of different baryonic components \citep{McKee15}.

\begin{figure}
    \centering
    \includegraphics[width=0.9\linewidth]{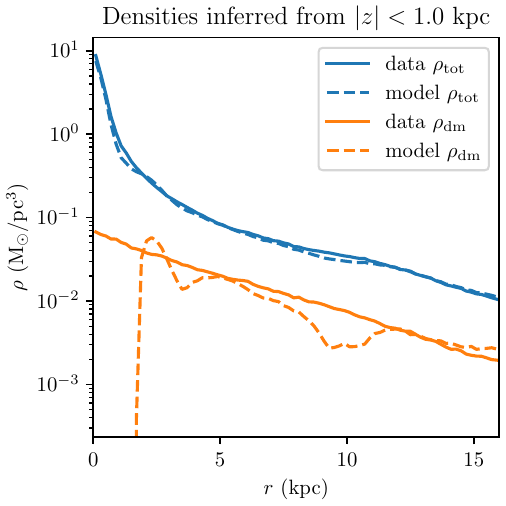}
    \caption{A comparison between the radial profiles for dark matter $\rho_\mathrm{dm}$ (orange lines) and overall matter $\rho_\mathrm{tot}$ (blue lines) predicted by the model (dashed lines) and the ground truth (solid lines, labeled ``data'') at time $t=\SI{2}{Gyr}$. The model estimate for the dark matter density is obtained by subtracting the true stellar density from the overall model density (See Section~\ref{sec:densities} for discussion). The profiles are obtained by calculating the average densities in a volume $|z| < \SI{1}{kpc}, r < \SI{16}{kpc}$ along $r$, where $r$ refers to the spherical radial distance. We accurately recover the radial profile of total (stellar plus dark-matter) density across the entire studied volume, which extends to \SI{16}{kpc}. In the central, heavily baryon-dominated region of the galaxy, even small fractional errors in the recovered total density are sufficient to significantly alter the recovered dark-matter density. However, for $r > \SI{2}{kpc}$, where the baryons are less dominant, we recover the dark-matter density profile to within a factor of two.}
    \label{fig:dm_t80}
\end{figure}

We provide a model estimate of the radial profile of the dark matter halo at time step $t=\SI{2}{Gyr}$ in Fig.~\ref{fig:dm_t80}. The modeled density deviates from the ground truth in volumes where baryonic matter dominates over dark matter, and where the overall density is small. For $r>\SI{2}{kpc}$, the dark matter profile is reconstructed to within a factor of two.

\begin{figure*}
    \centering
    \includegraphics[width=0.9\linewidth]{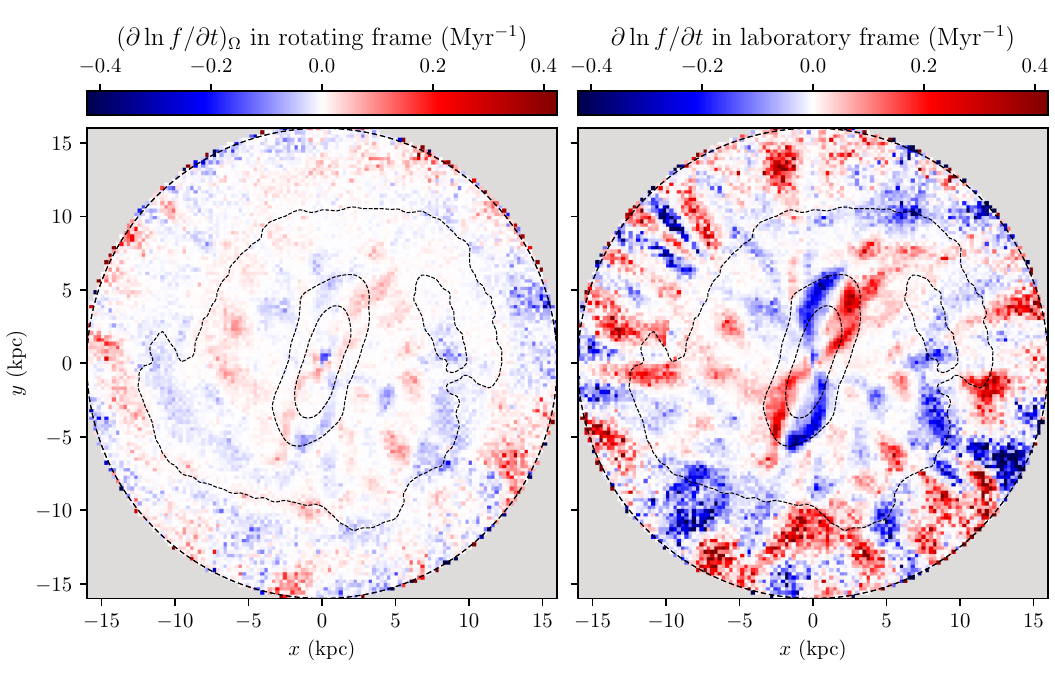}
    \caption{The level of inferred non-stationarity in the stellar population across the disc of our simulated galaxy, when applying \textit{Deep Potential} either in a rotating frame (left panel) or in the non-rotating ``laboratory'' frame (right panel). For time step $t=\SI{2}{Gyr}$, we calculate the non-stationarity, $\pdv*{\ln f}{t}$, calculated from our modeled distribution function and recovered potential. In the left panel, we work in the rotating frame inferred by \textit{Deep Potential} (with $\Omega = \SI{17.5}{km.s^{-1}.kpc^{-1}}$) and use the corresponding recovered gravitational potential. In the right panel, we work in the laboratory frame, and use the gravitational potential inferred by \textit{Deep Potential} for that frame. Both plots show the $x-y$ plane with $z=0$. For each bin in $x-y$, the median value of $\pdv*{\ln f}{t}$ is calculated by averaging over the velocity dimensions, weighted by the distribution function. $\pdv*{\ln f}{t}$ can be interpreted as the inverse timescale during which the distribution function undergoes significant changes. We overlay stellar isodensity contours, in order to indicate the location and extent of the galactic bar and spiral arms. When allowed to work in a rotating frame (left panel), \textit{Deep Potential} finds a gravitational potential that renders the system nearly stationary. When constrained to work in a non-rotating frame (right panel), \textit{Deep Potential} is unable to find a steady-state solution, and significant non-stationarities -- particularly along the galactic bar -- are present. This comparison demonstrates the improvement gained by allowing \textit{Deep Potential} to work in rotating frames, particularly in systems such as barred galaxies.
    }
    \label{fig:dfdt_x_y_t80}
\end{figure*}

\begin{table}
  \begin{tabular}{l|l|l|l}
    \hline
    $t$ & $\Omega_\mathrm{bar}$ & $\Omega_{2\mathrm{kpc}}$ (mirrored) & $\Omega_{4\mathrm{kpc}}$ (mirrored) \\
    (Gyr) & (\si{km.s^{-1}.kpc^{-1}}) & (\si{km.s^{-1}.kpc^{-1}}) & (\si{km.s^{-1}.kpc^{-1}})\\
    \hline
    2 & 20.7 & 23.7 (23.7) & 20.9 (19.6) \\
    4.25 & 12.15 & 14.1 (14.3) & 13.9 (13.1) \\
    9 & 8.1 & 9.81 (8.84) & 8.53 (7.96)
  \end{tabular}
  \caption{For all three time steps of our simulated galaxy, a comparison between the true rotation speed of the bar ($\Omega_{\rm bar}$) and the rotation speed that renders the system most stationary, according to \textit{Deep Potential}, in various sub-volumes imitating the Solar neighborhood. The sub-volumes are centered around a point at a distance of \SI{8}{kpc} and an angle of $\sim \ang{20}$ with respect to the galactic bar. We provide values for populations with radius \SI{2}{kpc} and \SI{4}{kpc}, and for their mirrored versions on the opposite side of the galaxy (values in parentheses). The bar rotation speed is captured to within $\sim 20\%$ and $\sim 15\%$ for the \SI{2}{kpc} and \SI{4}{kpc} datasets respectively.}
  \label{table:omegas_sun}
\end{table}

\subsection{Quantifying non-stationarities}
Real galaxies are never perfectly stationary. Even though we train the gravitational potential and $\Omega$ to minimize non-stationarities in the rotating frame, we do not expect them to be able perfectly render $(\pdv*{f}{t})_\Omega$ zero everywhere in phase space. This is also because the system is overconstrained: we are searching for a three-dimensional gravitational potential that renders the distribution function stationary at every point in six-dimensional phase space.

We can quantify non-stationarities by plotting the average $(\pdv*{\ln f}{t})_\Omega$ in different regions of phase space. $(\pdv*{\ln f}{t})_\Omega$ serves as a more useful measure than $(\pdv*{f}{t})_\Omega$ because it is more interpretable, corresponding to the inverse characteristic timescale during which the distribution function undergoes significant changes. Fig.~\ref{fig:dfdt_x_y_t80} shows a comparison between the non-stationarities of a gravitational potential trained in a rotating frame and one that is trained in the laboratory frame. We can see that in this system, the rotating frame yields significantly better results. Notably, the nonrotating potential has clear imprints from the central bar that are not visible in the rotating potential.

\begin{figure}
    \centering
    \includegraphics[width=0.9\linewidth]{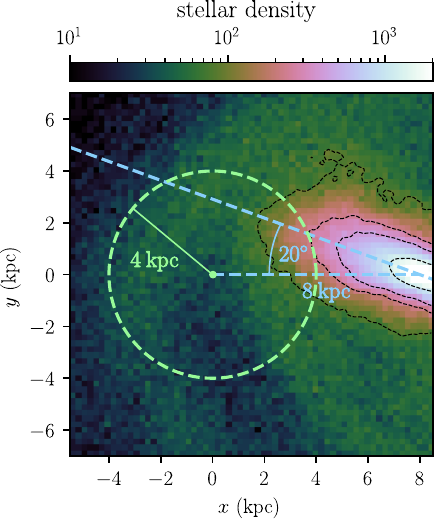}
    \caption{A visualization of a sub-volume of the simulated galaxy at $t=\SI{2}{Gyr}$, meant to imitate what one would see in the vicinity of the Sun in the Milky Way. The spherical sub-volume (denoted by a green dotted circle) is chosen to have a radius of \SI{4}{kpc}, and is centered around a point in the midplane, at a distance of \SI{8}{kpc} and angle of $\sim \ang{20}$ with respect to the major axis of the galactic bar (denoted by the diagonal blue dotted line), similar to how the Sun is positioned in the Milky Way. Using only stars in this sub-volume, \textit{Deep Potential} recovers the galactic bar rotation speed to within \SI{15}{\%}.}
    \label{fig:4kpc}
\end{figure}

\subsection{Rotation speed recovery in a sub-volume imitating the Solar neighborhood} \label{sec:solar_recovery}
This work so far has focused on the validation of \textit{Deep Potential} on the simulated galaxy as a whole. While this is important for the overall validation of the method within a rotating frame, we can also test how well the method recovers the bar rotation speed in a sub-volume which more closely resembles what we would observe from our position within the Milky Way. To this end, we select two spherical volumes of radius \SI{2}{kpc} and \SI{4}{kpc}, centered around a point at a distance of \SI{8}{kpc} and an angle of $\sim \ang{20}$ with respect to the galactic bar, akin to how the Sun is positioned in the Milky Way \citep{GaiaAxisymmetricDisc23}. Because of the two-fold symmetry of the galactic bar, we also produce mirrored datasets by reflecting the selected volumes with respect to the galactic center. We follow the same procedure for all time steps. One of the selected sub-volumes in the first time step is visualized in Fig.~\ref{fig:4kpc}.

At each time step, we train a normalizing flow and a gravitational potential, along with the rotation speed in the same fashion as outlined in the previous sections. \textit{Deep Potential} recovers the rotation speed to within \SI{20}{\%} and \SI{15}{\%} for the \SI{2}{kpc} and \SI{4}{kpc} datasets, respectively, for all time steps, which is, notably, slightly better than its performance when trained on the entire volume of the galaxy. We hypothesize that this may be due to the fact that the our sub-volume avoids the galactic center, where \textit{Deep Potential} has the greatest difficulty imposing stationarity (see the left panel of Fig.~\ref{fig:dfdt_x_y_t80}). In general, one can remain optimistic about the ability of \textit{Deep Potential} to recover the Milky Way bar rotation speed from stars observed with a few kiloparsecs of the Sun.

\section{Conclusions}
\label{sec:conclusions}

In this paper, we have demonstrated the \textit{Deep Potential} method in a self-consistent $N$-body simulation of a barred galaxy at three different time steps over the course of the evolution of the system. We have methodologically extended the \textit{Deep Potential} to impose stationarity in a rotating frame. We achieve this by selecting a population of stellar particles and simultaneously fitting a gravitational potential and a rotation speed that best renders the population stationary in the rotating frame. For our $N$-body simulation, in a \SI{16}{kpc} dataset encompassing the majority of the simulated galaxy, \textit{Deep Potential} recovers the rotation speed of the galactic bar to within \SI{20}{\%}, accelerations in the system to within \SI{20}{\%} and densities to within \SI{50}{\%}. We also recover the radial dark matter profile in outer regions of the galaxy ($r>\SI{2}{kpc}$). The main features of the galaxy, such as the spiral arms and the central bar, are successfully recovered, even in the presence of strong intrinsic non-stationarities in the galactic stellar population. The modeled densities, however, exhibit small scale fluctuations, which could be caused by the intrinsic non-stationarity of the data, the smoothing length of the particles, or discrepancies in the modeling of the underlying distribution function with a normalizing flow. We have additionally demonstrated that \textit{Deep Potential} is capable of recovering the rotation speed in smaller sub-volumes imitating the Solar neighborhood: the rotation speed of the bar is captured to within \SI{20}{\%} and \SI{15}{\%} for sub-volumes of radius \SI{2}{kpc} and \SI{4}{kpc} respectively.

Working in an arbitrarily rotating frame is important for modeling real-life galaxies, which often have rotating features such as spiral arms or a central bar. The Milky Way serves as a good candidate for the next application of \textit{Deep Potential}, with the availability of six-dimensional phase space info for tens of millions of stars from Gaia Data Release~3 \citep{GaiaDR323}. In particular, \textit{Deep Potential} may be able to determine the pattern speed of the Milky Way bar, as well as the radial dark-matter density profile. Indeed, there has been recent work on applying normalizing-flow-based modeling on the Milky Way. \citet{Lim2023} saw the first application on the Milky Way for inferring local dark matter densities.

There are also several other avenues for future development. One could extend the \textit{Deep Potential} formalism to account for non-uniform observational completeness in the kinematic tracers and errors on measured phase-space locations. Further, if there is data available for additional dimensions that are relevant for selecting populations which share the same dynamical history, such as metallicity or alpha abundance, one could train normalizing flows incorporating those dimensions, and then enforce stationarity on the distribution function while conditioning on the extra dimensions. Finally, one can model systems where full six-dimensional phase-space data is not available by imposing symmetries on the system, such as axisymmetry. While this is not relevant for the Milky Way, it can be important for external galaxies or compact systems where some of the dimensions (such as distance) are missing.

\section*{Acknowledgements}
This work was supported by funding from the Alexander von Humboldt Foundation, through Gregory M. Green's Sofja Kovalevskaja Award, and made use of the HPC system Raven at the Max Planck Computing and Data Facility. The authors thank Paola Di Matteo for providing the simulation data, Hans-Walter Rix for useful discussions about adapting \textit{Deep Potential} to a rotating frame, and Tristan Cantat-Gaudin for suggesting that we test our method in a Solar-like galactic sub-volume. This work has made use of the computational resources obtained through the DARI grant A0120410154 (P.I. : P. Di Matteo).

\section*{Data Availability}

All of our code, trained models, training data, and simulation snapshots, as well as Python notebooks to generate paper plots, are publicly available under a permissive license that allows reuse and modification with attribution, both in archived form at \url{https://doi.org/10.5281/zenodo.8390759} and in active development at \url{https://github.com/gregreen/deep-potential}.



\bibliographystyle{mnras}
\bibliography{refs}

\begin{thebibliography}{}
\makeatletter
\relax
\def\mn@urlcharsother{\let\do\@makeother \do\$\do\&\do\#\do\^\do\_\do\%\do\~}
\def\mn@doi{\begingroup\mn@urlcharsother \@ifnextchar [ {\mn@doi@}
  {\mn@doi@[]}}
\def\mn@doi@[#1]#2{\def\@tempa{#1}\ifx\@tempa\@empty \href
  {http://dx.doi.org/#2} {doi:#2}\else \href {http://dx.doi.org/#2} {#1}\fi
  \endgroup}
\def\mn@eprint#1#2{\mn@eprint@#1:#2::\@nil}
\def\mn@eprint@arXiv#1{\href {http://arxiv.org/abs/#1} {{\tt arXiv:#1}}}
\def\mn@eprint@dblp#1{\href {http://dblp.uni-trier.de/rec/bibtex/#1.xml}
  {dblp:#1}}
\def\mn@eprint@#1:#2:#3:#4\@nil{\def\@tempa {#1}\def\@tempb {#2}\def\@tempc
  {#3}\ifx \@tempc \@empty \let \@tempc \@tempb \let \@tempb \@tempa \fi \ifx
  \@tempb \@empty \def\@tempb {arXiv}\fi \@ifundefined
  {mn@eprint@\@tempb}{\@tempb:\@tempc}{\expandafter \expandafter \csname
  mn@eprint@\@tempb\endcsname \expandafter{\@tempc}}}

\bibitem[\protect\citeauthoryear{Abadi et~al.,}{Abadi
  et~al.}{2015}]{tensorflow2015-whitepaper}
Abadi M.,  et~al., 2015, {TensorFlow}: Large-Scale Machine Learning on
  Heterogeneous Systems, \url {https://www.tensorflow.org/}

\bibitem[\protect\citeauthoryear{{An}, {Naik}, {Evans}  \& {Burrage}}{{An}
  et~al.}{2021}]{An2021}
{An} J.,  {Naik} A.~P.,  {Evans} N.~W.,   {Burrage} C.,  2021, \mn@doi [\mnras]
  {10.1093/mnras/stab2049}, \href
  {https://ui.adsabs.harvard.edu/abs/2021MNRAS.506.5721A} {506, 5721}

\bibitem[\protect\citeauthoryear{{Barnes} \& {Hut}}{{Barnes} \&
  {Hut}}{1986}]{BarnesandHut1986}
{Barnes} J.,  {Hut} P.,  1986, \mn@doi [\nat] {10.1038/324446a0}, \href
  {https://ui.adsabs.harvard.edu/abs/1986Natur.324..446B} {324, 446}

\bibitem[\protect\citeauthoryear{{Binney} \& {Tremaine}}{{Binney} \&
  {Tremaine}}{2008}]{BinneyTremaine2008}
{Binney} J.,  {Tremaine} S.,  2008, {Galactic Dynamics: Second Edition}.
Princeton University Press

\bibitem[\protect\citeauthoryear{{Binney}, {Gerhard}, {Stark}, {Bally}  \&
  {Uchida}}{{Binney} et~al.}{1991}]{Binneyetal1991}
{Binney} J.,  {Gerhard} O.~E.,  {Stark} A.~A.,  {Bally} J.,   {Uchida} K.~I.,
  1991, \mn@doi [\mnras] {10.1093/mnras/252.2.210}, \href
  {https://ui.adsabs.harvard.edu/abs/1991MNRAS.252..210B} {252, 210}

\bibitem[\protect\citeauthoryear{{Binney}, {Gerhard}  \& {Spergel}}{{Binney}
  et~al.}{1997}]{Binneyetal1997}
{Binney} J.,  {Gerhard} O.,   {Spergel} D.,  1997, \mn@doi [\mnras]
  {10.1093/mnras/288.2.365}, \href
  {https://ui.adsabs.harvard.edu/abs/1997MNRAS.288..365B} {288, 365}

\bibitem[\protect\citeauthoryear{{Blitz} \& {Spergel}}{{Blitz} \&
  {Spergel}}{1991}]{BlitzandSpergel1991}
{Blitz} L.,  {Spergel} D.~N.,  1991, \mn@doi [\apj] {10.1086/170535}, \href
  {https://ui.adsabs.harvard.edu/abs/1991ApJ...379..631B} {379, 631}

\bibitem[\protect\citeauthoryear{{Bovy} \& {Rix}}{{Bovy} \&
  {Rix}}{2013}]{Bovy13}
{Bovy} J.,  {Rix} H.-W.,  2013, \mn@doi [\apj] {10.1088/0004-637X/779/2/115},
  \href {https://ui.adsabs.harvard.edu/abs/2013ApJ...779..115B} {779, 115}

\bibitem[\protect\citeauthoryear{Bradbury et~al.,}{Bradbury
  et~al.}{2018}]{jax2018github}
Bradbury J.,  et~al., 2018, {JAX}: composable transformations of
  {P}ython+{N}um{P}y programs, \url {http://github.com/google/jax}

\bibitem[\protect\citeauthoryear{{Buckley}, {Lim}, {Putney}  \&
  {Shih}}{{Buckley} et~al.}{2023a}]{Buckley2022GalacticDM}
{Buckley} M.~R.,  {Lim} S.~H.,  {Putney} E.,   {Shih} D.,  2023a, \mn@doi
  [\mnras] {10.1093/mnras/stad843}, \href
  {https://ui.adsabs.harvard.edu/abs/2023MNRAS.521.5100B} {521, 5100}

\bibitem[\protect\citeauthoryear{{Buckley}, {Lim}, {Putney}  \&
  {Shih}}{{Buckley} et~al.}{2023b}]{Buckley23}
{Buckley} M.~R.,  {Lim} S.~H.,  {Putney} E.,   {Shih} D.,  2023b, \mn@doi
  [\mnras] {10.1093/mnras/stad843}, \href
  {https://ui.adsabs.harvard.edu/abs/2023MNRAS.521.5100B} {521, 5100}

\bibitem[\protect\citeauthoryear{{Chakrabarti}, {Chang}, {Lam}, {Vigeland}  \&
  {Quillen}}{{Chakrabarti} et~al.}{2021}]{Chakrabarti21}
{Chakrabarti} S.,  {Chang} P.,  {Lam} M.~T.,  {Vigeland} S.~J.,   {Quillen}
  A.~C.,  2021, \mn@doi [\apjl] {10.3847/2041-8213/abd635}, \href
  {https://ui.adsabs.harvard.edu/abs/2021ApJ...907L..26C} {907, L26}

\bibitem[\protect\citeauthoryear{{Churchwell} et~al.,}{{Churchwell}
  et~al.}{2009}]{Churchweletal2009}
{Churchwell} E.,  et~al., 2009, \mn@doi [\pasp] {10.1086/597811}, \href
  {https://ui.adsabs.harvard.edu/abs/2009PASP..121..213C} {121, 213}

\bibitem[\protect\citeauthoryear{{Clarke} \& {Gerhard}}{{Clarke} \&
  {Gerhard}}{2022}]{Clarke22}
{Clarke} J.~P.,  {Gerhard} O.,  2022, \mn@doi [\mnras] {10.1093/mnras/stac603},
  \href {https://ui.adsabs.harvard.edu/abs/2022MNRAS.512.2171C} {512, 2171}

\bibitem[\protect\citeauthoryear{{Dillon} et~al.,}{{Dillon}
  et~al.}{2017}]{Dillon17}
{Dillon} J.~V.,  et~al., 2017, \mn@doi [arXiv e-prints]
  {10.48550/arXiv.1711.10604}, \href
  {https://ui.adsabs.harvard.edu/abs/2017arXiv171110604D} {p. arXiv:1711.10604}

\bibitem[\protect\citeauthoryear{{Finlay}, {Jacobsen}, {Nurbekyan}  \&
  {Oberman}}{{Finlay} et~al.}{2020}]{Finlay2020}
{Finlay} C.,  {Jacobsen} J.-H.,  {Nurbekyan} L.,   {Oberman} A.~M.,  2020,
  \mn@doi [arXiv e-prints] {10.48550/arXiv.2002.02798}, \href
  {https://ui.adsabs.harvard.edu/abs/2020arXiv200202798F} {p. arXiv:2002.02798}

\bibitem[\protect\citeauthoryear{{Fragkoudi}, {Di Matteo}, {Haywood},
  {G{\'o}mez}, {Combes}, {Katz}  \& {Semelin}}{{Fragkoudi}
  et~al.}{2017}]{Fragkoudietal2017}
{Fragkoudi} F.,  {Di Matteo} P.,  {Haywood} M.,  {G{\'o}mez} A.,  {Combes} F.,
  {Katz} D.,   {Semelin} B.,  2017, \mn@doi [\aap]
  {10.1051/0004-6361/201630244}, \href
  {https://ui.adsabs.harvard.edu/abs/2017A&A...606A..47F} {606, A47}

\bibitem[\protect\citeauthoryear{{Gaia Collaboration} et~al.,}{{Gaia
  Collaboration} et~al.}{2016}]{Gaia16}
{Gaia Collaboration} et~al., 2016, \mn@doi [\aap]
  {10.1051/0004-6361/201629272}, \href
  {https://ui.adsabs.harvard.edu/abs/2016A&A...595A...1G} {595, A1}

\bibitem[\protect\citeauthoryear{{Gaia Collaboration} et~al.,}{{Gaia
  Collaboration} et~al.}{2023a}]{GaiaDR323}
{Gaia Collaboration} et~al., 2023a, \mn@doi [\aap]
  {10.1051/0004-6361/202243940}, \href
  {https://ui.adsabs.harvard.edu/abs/2023A&A...674A...1G} {674, A1}

\bibitem[\protect\citeauthoryear{{Gaia Collaboration} et~al.,}{{Gaia
  Collaboration} et~al.}{2023b}]{GaiaAxisymmetricDisc23}
{Gaia Collaboration} et~al., 2023b, \mn@doi [\aap]
  {10.1051/0004-6361/202243797}, \href
  {https://ui.adsabs.harvard.edu/abs/2023A&A...674A..37G} {674, A37}

\bibitem[\protect\citeauthoryear{{Georgelin} \& {Georgelin}}{{Georgelin} \&
  {Georgelin}}{1976}]{GeorgelinandGeorgelin1976}
{Georgelin} Y.~M.,  {Georgelin} Y.~P.,  1976, \aap, \href
  {https://ui.adsabs.harvard.edu/abs/1976A&A....49...57G} {49, 57}

\bibitem[\protect\citeauthoryear{{Gerhard}}{{Gerhard}}{2002}]{Gerhard2002}
{Gerhard} O.,  2002, in {Da Costa} G.~S.,  {Sadler} E.~M.,   {Jerjen} H.,  eds,
   Astronomical Society of the Pacific Conference Series Vol. 273, The
  Dynamics, Structure \& History of Galaxies: A Workshop in Honour of Professor
  Ken Freeman. p.~73 (\mn@eprint {arXiv} {astro-ph/0203109})

\bibitem[\protect\citeauthoryear{{Ghez}, {Morris}, {Becklin}, {Tanner}  \&
  {Kremenek}}{{Ghez} et~al.}{2000}]{Ghez00}
{Ghez} A.~M.,  {Morris} M.,  {Becklin} E.~E.,  {Tanner} A.,   {Kremenek} T.,
  2000, \mn@doi [\nat] {10.1038/35030032}, \href
  {https://ui.adsabs.harvard.edu/abs/2000Natur.407..349G} {407, 349}

\bibitem[\protect\citeauthoryear{{Ghosh} \& {Di Matteo}}{{Ghosh} \& {Di
  Matteo}}{2023}]{GhoshDiMatteo2023}
{Ghosh} S.,  {Di Matteo} P.,  2023, \mn@doi [arXiv e-prints]
  {10.48550/arXiv.2308.10948}, \href
  {https://ui.adsabs.harvard.edu/abs/2023arXiv230810948G} {p. arXiv:2308.10948}

\bibitem[\protect\citeauthoryear{{Ghosh}, {Trick}  \& {Green}}{{Ghosh}
  et~al.}{2023a}]{Ghosh+Roadmap2023}
{Ghosh} S.,  {Trick} W.~H.,   {Green} G.~M.,  2023a, \mn@doi [\mnras]
  {10.1093/mnras/stad1525}, \href
  {https://ui.adsabs.harvard.edu/abs/2023MNRAS.523..991G} {523, 991}

\bibitem[\protect\citeauthoryear{{Ghosh}, {Fragkoudi}, {Di Matteo}  \&
  {Saha}}{{Ghosh} et~al.}{2023b}]{Ghoshetal2022}
{Ghosh} S.,  {Fragkoudi} F.,  {Di Matteo} P.,   {Saha} K.,  2023b, \mn@doi
  [\aap] {10.1051/0004-6361/202245275}, \href
  {https://ui.adsabs.harvard.edu/abs/2023A&A...674A.128G} {674, A128}

\bibitem[\protect\citeauthoryear{{Grathwohl}, {Chen}, {Bettencourt},
  {Sutskever}  \& {Duvenaud}}{{Grathwohl} et~al.}{2018}]{Grathwohl18}
{Grathwohl} W.,  {Chen} R. T.~Q.,  {Bettencourt} J.,  {Sutskever} I.,
  {Duvenaud} D.,  2018, \mn@doi [arXiv e-prints] {10.48550/arXiv.1810.01367},
  \href {https://ui.adsabs.harvard.edu/abs/2018arXiv181001367G} {p.
  arXiv:1810.01367}

\bibitem[\protect\citeauthoryear{{Green} \& {Ting}}{{Green} \&
  {Ting}}{2020}]{Green20}
{Green} G.~M.,  {Ting} Y.-S.,  2020, \mn@doi [arXiv e-prints]
  {10.48550/arXiv.2011.04673}, \href
  {https://ui.adsabs.harvard.edu/abs/2020arXiv201104673G} {p. arXiv:2011.04673}

\bibitem[\protect\citeauthoryear{{Green}, {Ting}  \& {Kamdar}}{{Green}
  et~al.}{2023}]{Green23}
{Green} G.~M.,  {Ting} Y.-S.,   {Kamdar} H.,  2023, \mn@doi [\apj]
  {10.3847/1538-4357/aca3a7}, \href
  {https://ui.adsabs.harvard.edu/abs/2023ApJ...942...26G} {942, 26}

\bibitem[\protect\citeauthoryear{{Hammersley}, {Garz{\'o}n}, {Mahoney},
  {L{\'o}pez-Corredoira}  \& {Torres}}{{Hammersley}
  et~al.}{2000}]{Hammersleyetal2000}
{Hammersley} P.~L.,  {Garz{\'o}n} F.,  {Mahoney} T.~J.,  {L{\'o}pez-Corredoira}
  M.,   {Torres} M.~A.~P.,  2000, \mn@doi [\mnras]
  {10.1046/j.1365-8711.2000.03858.x}, \href
  {https://ui.adsabs.harvard.edu/abs/2000MNRAS.317L..45H} {317, L45}

\bibitem[\protect\citeauthoryear{{Kobyzev}, {Prince}  \& {Brubaker}}{{Kobyzev}
  et~al.}{2019}]{Kobyzev2019}
{Kobyzev} I.,  {Prince} S. J.~D.,   {Brubaker} M.~A.,  2019, \mn@doi [arXiv
  e-prints] {10.48550/arXiv.1908.09257}, \href
  {https://ui.adsabs.harvard.edu/abs/2019arXiv190809257K} {p. arXiv:1908.09257}

\bibitem[\protect\citeauthoryear{{Lim}, {Putney}, {Buckley}  \& {Shih}}{{Lim}
  et~al.}{2023}]{Lim2023}
{Lim} S.~H.,  {Putney} E.,  {Buckley} M.~R.,   {Shih} D.,  2023, \mn@doi [arXiv
  e-prints] {10.48550/arXiv.2305.13358}, \href
  {https://ui.adsabs.harvard.edu/abs/2023arXiv230513358L} {p. arXiv:2305.13358}

\bibitem[\protect\citeauthoryear{{Liszt} \& {Burton}}{{Liszt} \&
  {Burton}}{1980}]{LisztandBurton1980}
{Liszt} H.~S.,  {Burton} W.~B.,  1980, \mn@doi [\apj] {10.1086/157803}, \href
  {https://ui.adsabs.harvard.edu/abs/1980ApJ...236..779L} {236, 779}

\bibitem[\protect\citeauthoryear{{Liu}, {Jiang}, {He}, {Chen}, {Liu}, {Gao}  \&
  {Han}}{{Liu} et~al.}{2019}]{Liu2019}
{Liu} L.,  {Jiang} H.,  {He} P.,  {Chen} W.,  {Liu} X.,  {Gao} J.,   {Han} J.,
  2019, \mn@doi [arXiv e-prints] {10.48550/arXiv.1908.03265}, \href
  {https://ui.adsabs.harvard.edu/abs/2019arXiv190803265L} {p. arXiv:1908.03265}

\bibitem[\protect\citeauthoryear{{Magorrian}}{{Magorrian}}{2014}]{Magorrian14}
{Magorrian} J.,  2014, \mn@doi [\mnras] {10.1093/mnras/stt2031}, \href
  {https://ui.adsabs.harvard.edu/abs/2014MNRAS.437.2230M} {437, 2230}

\bibitem[\protect\citeauthoryear{{McKee}, {Parravano}  \& {Hollenbach}}{{McKee}
  et~al.}{2015}]{McKee15}
{McKee} C.~F.,  {Parravano} A.,   {Hollenbach} D.~J.,  2015, \mn@doi [\apj]
  {10.1088/0004-637X/814/1/13}, \href
  {https://ui.adsabs.harvard.edu/abs/2015ApJ...814...13M} {814, 13}

\bibitem[\protect\citeauthoryear{{McMillan} \& {Binney}}{{McMillan} \&
  {Binney}}{2008}]{McMillan08}
{McMillan} P.~J.,  {Binney} J.~J.,  2008, \mn@doi [\mnras]
  {10.1111/j.1365-2966.2008.13767.x}, \href
  {https://ui.adsabs.harvard.edu/abs/2008MNRAS.390..429M} {390, 429}

\bibitem[\protect\citeauthoryear{{Miyamoto} \& {Nagai}}{{Miyamoto} \&
  {Nagai}}{1975}]{MiyamatoandNagai1975}
{Miyamoto} M.,  {Nagai} R.,  1975, \pasj, \href
  {https://ui.adsabs.harvard.edu/abs/1975PASJ...27..533M} {27, 533}

\bibitem[\protect\citeauthoryear{{Naik}, {An}, {Burrage}  \& {Evans}}{{Naik}
  et~al.}{2022}]{Naik2022GalacticAcc}
{Naik} A.~P.,  {An} J.,  {Burrage} C.,   {Evans} N.~W.,  2022, \mn@doi [\mnras]
  {10.1093/mnras/stac153}, \href
  {https://ui.adsabs.harvard.edu/abs/2022MNRAS.511.1609N} {511, 1609}

\bibitem[\protect\citeauthoryear{{Oort}, {Kerr}  \& {Westerhout}}{{Oort}
  et~al.}{1958}]{Oort1958}
{Oort} J.~H.,  {Kerr} F.~J.,   {Westerhout} G.,  1958, \mn@doi [\mnras]
  {10.1093/mnras/118.4.379}, \href
  {https://ui.adsabs.harvard.edu/abs/1958MNRAS.118..379O} {118, 379}

\bibitem[\protect\citeauthoryear{{Paszke} et~al.,}{{Paszke}
  et~al.}{2019}]{NEURIPS2019_9015}
{Paszke} A.,  et~al., 2019, \mn@doi [arXiv e-prints]
  {10.48550/arXiv.1912.01703}, \href
  {https://ui.adsabs.harvard.edu/abs/2019arXiv191201703P} {p. arXiv:1912.01703}

\bibitem[\protect\citeauthoryear{Phillips, Ravi, Ebadi  \& Walsworth}{Phillips
  et~al.}{2021}]{Phillips21}
Phillips D.~F.,  Ravi A.,  Ebadi R.,   Walsworth R.~L.,  2021, \mn@doi [Phys.
  Rev. Lett.] {10.1103/PhysRevLett.126.141103}, 126, 141103

\bibitem[\protect\citeauthoryear{{Plummer}}{{Plummer}}{1911}]{Plummer1911}
{Plummer} H.~C.,  1911, \mn@doi [\mnras] {10.1093/mnras/71.5.460}, \href
  {https://ui.adsabs.harvard.edu/abs/1911MNRAS..71..460P} {71, 460}

\bibitem[\protect\citeauthoryear{{Press}, {Flannery}  \& {Teukolsky}}{{Press}
  et~al.}{1986}]{Pressetal1986}
{Press} W.~H.,  {Flannery} B.~P.,   {Teukolsky} S.~A.,  1986, {Numerical
  recipes. The art of scientific computing}

\bibitem[\protect\citeauthoryear{{Reid} et~al.,}{{Reid}
  et~al.}{2014}]{Reidetal2014}
{Reid} M.~J.,  et~al., 2014, \mn@doi [\apj] {10.1088/0004-637X/783/2/130},
  \href {https://ui.adsabs.harvard.edu/abs/2014ApJ...783..130R} {783, 130}

\bibitem[\protect\citeauthoryear{{Rodionov}, {Athanassoula}  \&
  {Sotnikova}}{{Rodionov} et~al.}{2009}]{Rodionovetal2009}
{Rodionov} S.~A.,  {Athanassoula} E.,   {Sotnikova} N.~Y.,  2009, \mn@doi
  [\mnras] {10.1111/j.1365-2966.2008.14110.x}, \href
  {https://ui.adsabs.harvard.edu/abs/2009MNRAS.392..904R} {392, 904}

\bibitem[\protect\citeauthoryear{{Schwarzschild}}{{Schwarzschild}}{1979}]{Schwarzschild79}
{Schwarzschild} M.,  1979, \mn@doi [\apj] {10.1086/157282}, \href
  {https://ui.adsabs.harvard.edu/abs/1979ApJ...232..236S} {232, 236}

\bibitem[\protect\citeauthoryear{{Semelin} \& {Combes}}{{Semelin} \&
  {Combes}}{2002}]{SemelinandCombes2002}
{Semelin} B.,  {Combes} F.,  2002, \mn@doi [\aap] {10.1051/0004-6361:20020547},
  \href {https://ui.adsabs.harvard.edu/abs/2002A&A...388..826S} {388, 826}

\bibitem[\protect\citeauthoryear{{Shen} \& {Zheng}}{{Shen} \&
  {Zheng}}{2020}]{Shen20}
{Shen} J.,  {Zheng} X.-W.,  2020, \mn@doi [Research in Astronomy and
  Astrophysics] {10.1088/1674-4527/20/10/159}, \href
  {https://ui.adsabs.harvard.edu/abs/2020RAA....20..159S} {20, 159}

\bibitem[\protect\citeauthoryear{Silverwood \& Easther}{Silverwood \&
  Easther}{2019}]{silverwood_easther_2019}
Silverwood H.,  Easther R.,  2019, \mn@doi [Publications of the Astronomical
  Society of Australia] {10.1017/pasa.2019.25}, 36, e038

\bibitem[\protect\citeauthoryear{{Syer} \& {Tremaine}}{{Syer} \&
  {Tremaine}}{1996}]{Syer96}
{Syer} D.,  {Tremaine} S.,  1996, \mn@doi [\mnras] {10.1093/mnras/282.1.223},
  \href {https://ui.adsabs.harvard.edu/abs/1996MNRAS.282..223S} {282, 223}

\bibitem[\protect\citeauthoryear{{Wegg} \& {Gerhard}}{{Wegg} \&
  {Gerhard}}{2013}]{WegandGerhard2013}
{Wegg} C.,  {Gerhard} O.,  2013, \mn@doi [\mnras] {10.1093/mnras/stt1376},
  \href {https://ui.adsabs.harvard.edu/abs/2013MNRAS.435.1874W} {435, 1874}

\bibitem[\protect\citeauthoryear{{Weinberg}}{{Weinberg}}{1992}]{Weinberg1992}
{Weinberg} M.~D.,  1992, \mn@doi [\apj] {10.1086/170853}, \href
  {https://ui.adsabs.harvard.edu/abs/1992ApJ...384...81W} {384, 81}

\makeatother
\end{thebibliography}



\appendix

\section{Deriving the stationarity condition in a rotating frame}
\label{app:stationarity}

We wish to impose stationarity in a rotating frame. We describe the lab frame using the coordinates $\left(t,\vec{x},\vec{v}\right)$, and the rotating frame using primed coordinates, $\left(t',\vec{x}',\vec{v}'\right)$. These two coordinate systems are related by the transformation
\begin{align}
  t' &= t ,
    &&
    & t &= t' , \\
  \vec{x}' &= R\left(\Omega \, t\right)\vec{x} ,
    &\Longleftrightarrow&
    & \vec{x} &= R\left(-\Omega \, t'\right)\vec{x}' , \\
  \vec{v}' &= R\left(\Omega \, t\right)\vec{v} ,
    &&
    & \vec{v} &= R\left(-\Omega \, t'\right)\vec{v}' ,
\end{align}
where $R\left(\theta\right)$ is a rotation matrix (through angle $\theta$ about some particular axis), and $\Omega$ is the angular frequency of the rotation. We will need to take derivatives of the rotation matrix, and for that, it will be useful to express $R$ in exponential form:
\begin{align}
  R\left(\theta\right) = e^{\theta L} ,
\end{align}
where $L$ is a rotation-generating matrix. The indices of $L$ can be related to the scalar angular frequency, $\Omega$, and the vector angular frequency, $\vec{\Omega}$, by
\begin{align}
  \Omega L_{ij} = \epsilon_{ijk} \Omega_k ,
  \label{eqn:L-omega-relation}
\end{align}
where $\epsilon_{ijk}$ is the Levi-Civita symbol. The derivative of $R$ can now be calculated as
\begin{align}
  \dv{R\left(\theta\right)}{\theta}
  &= \dv{}{\theta} e^{\theta L}
  = L e^{\theta L}
  = L R\left(\theta\right) \, .
  \label{eqn:dR_dtheta}
\end{align}
We impose stationarity on the distribution function in the rotating frame: $\pdv*{f}{t'} = 0$. However, we wish to express this condition in terms of lab-frame (unprimed) coordinates:
\begin{align}
  0 = \pdv{f}{t^{\prime}}
  &=
      \pdv{f}{t}\pdv{t}{t'}
    + \pdv{f}{\vec{x}}\cdot\pdv{\vec{x}}{t'}
    + \pdv{f}{\vec{v}}\cdot\pdv{\vec{v}}{t'}
  \, .
  \label{eqn:stationarity-pdv-conversion}
\end{align}
The derivative $\pdv*{t}{t'} = 1$ is trivial, but the derivatives $\pdv*{\vec{x}}{t'}$ and $\pdv*{\vec{v}}{t'}$ require some calculation:
\begin{align}
  \pdv{\vec{x}}{t'}
  &= \dv{R\left(-\Omega t'\right)}{t'} \vec{x}'
  = \dv{R\left(-\Omega t'\right)}{t'} R\left(\Omega t\right)\vec{x} .
\end{align}
Using Eq.~\eqref{eqn:dR_dtheta}, we find
\begin{align}
  \pdv{\vec{x}}{t'}
  &= -\omega L R\left(-\Omega t\right) R\left(\Omega t\right) \vec{x}
  = -\Omega L \vec{x} \, .
\end{align}
Writing this equation in index notation and substituting in Eq.~\eqref{eqn:L-omega-relation}, we obtain
\begin{align}
  \pdv{x_i}{t'} = -\epsilon_{ijk} \Omega_k x_j \, .
\end{align}
In vector notation, this is equivalent to
\begin{align}
  \pdv{\vec{x}}{t'} = \vec{\Omega} \times \vec{x} \, .
\end{align}
There is an identical result for $\vec{v}$. Plugging the above into Eq.~\ref{eqn:stationarity-pdv-conversion}, we obtain the generalized stationarity condition in terms of lab-frame (unprimed) coordinates:
\begin{align}
  0 = \pdv{f}{t'}
    = \pdv{f}{t}
      + \big(\vec{\Omega}\times\vec{x}\big) \cdot \pdv{f}{\vec{x}}
      + \big(\vec{\Omega}\times\vec{v}\big) \cdot \pdv{f}{\vec{v}}
  .
  \label{eqn:stationarity-lab-frame}
\end{align}
We can use the same approach to determine the stationarity condition for a moving (but non-rotating) frame with a positional offset:
\begin{align}
  t' &= t , \\
  \vec{x}' &= \vec{x} - \vec{x}_0 - \vec{v}_0 t , \\
  \vec{v}' &= \vec{v} - \vec{v}_0 .
\end{align}
The partial derivatives w.r.t. time in the moving (primed) and lab (unprimed) frame are related by
\begin{align}
  \pdv{f}{t'} = \pdv{f}{t} + \vec{v}_0 \cdot \pdv{f}{\vec{x}} \, .
\end{align}
We can chain the transformations for moving and rotating frames together. In the Milky Way, the most natural frame in which to impose stationarity is a frame centered on the Galactic Center, which is rotating about the $z$-axis (either with the bar or with the spiral arms). Denote the lab frame by unprimed coordinates, the Galactic frame by primed coordinates, and the rotating frame by double-primed coordinates. At time $t = 0$, the stationarity condition is given by
\begin{align}
  0 &= \pdv{f}{t''} \\
  &= \pdv{f}{t'}
    + \left(\vec{\Omega}\times\vec{x}'\right) \cdot \pdv{f}{\vec{x}'}
    + \left(\vec{\Omega}\times\vec{v}'\right) \cdot \pdv{f}{\vec{v}'}
    \\
  &= \pdv{f}{t}
    + \left[\vec{\Omega}\times\left(\vec{x}-\vec{x}_0\right)+\vec{v}_0\right] \cdot \pdv{f}{\vec{x}}
    + \left[\vec{\Omega}\times\left(\vec{v}-\vec{v}_0\right)\right] \cdot \pdv{f}{\vec{v}}
  ,
\end{align}
where we have used the fact that $\pdv{f}{\vec{x}'} = \pdv{f}{\vec{x}}$ and $\pdv{f}{\vec{v}'} = \pdv{f}{\vec{v}}$. Defining
\begin{align}
  \vec{u}\left(\vec{x}\right) &\equiv \vec{\Omega}\times\left(\vec{x}-\vec{x}_0\right) + \vec{v}_0 , \\
  \vec{w}\left(\vec{v}\right) &\equiv \vec{\Omega}\times\left(\vec{v}-\vec{v}_0\right) ,
\end{align}
the stationarity condition is then given by
\begin{align}
  0 = \pdv{f}{t}
      + \vec{u}\left(\vec{x}\right)\cdot\pdv{f}{\vec{x}}
      + \vec{w}\left(\vec{v}\right)\cdot\pdv{f}{\vec{v}}
  \, ,
\end{align}
as in Eq.~\eqref{eqn:generalized-stationarity}. In the body of the text, $\pdv*{f}{t''}$ is denoted by $\left(\pdv*{f}{t}\right)_{\Omega}$.

\section{Normalizing flow padding}
\label{app:df_padding}

In Section~\ref{sec:constructing_df}, we discuss how modeling an individual time step involves separating the dataset into four separate shells and adding ``virtual'' particles outside the boundaries of each shell. The virtual particles form a padding around the shell, which transforms the sharp density cutoff into a smooth roll-off. Each shell has an inner cylindrical boundary $\rin$, an outer cylindrical boundary $\rout$, and two faces at $z=\pm H$. The values of the inner and outer boundaries are specified in Table~\ref{table:shells}. In practice, we find that it is enough to match the derivatives up to second order between the shell and the padding. The characteristic padding scale is chosen to be $\sim \SI{20}{\%}$ of the scale of the shell along the respective spatial dimension.

We set the density distribution of virtual particles to connect smoothly to the density of real particles inside the ``observed'' volume, such that there is no spatial discontinuity in the total density of particles (virtual plus real). One possible approach would be to generate the virtual particles by sampling from a distribution that smoothly extends the population of observed particles past the chosen volume. In order to do this, one would have to create a distribution of virtual particles whose density and density derivatives (up to a chosen order) match those of the observed particles at the boundaries of the chosen volume. This is, however, a difficult problem, which is nearly identical to the original problem of learning the normalizing flow of the observed particles.

Instead, we opt for the approach of selecting the virtual particles from the set of all stellar particles, including those beyond the modeled volume, as that can generally be assumed to be a spatially smooth population. We realize this by selecting each star to belong to the padded dataset with some probability $p(\vec x)$ that is constant inside the selected volume, and then drops off in a sufficiently smooth manner outside the selected volume.

In our cylindrical volumes, we define $p(\vec x)$ as follows:
\begin{equation}
p(\vec x) = p_0 g_R(R)g_z(z)h(r),
\end{equation}
where $R$ is the cylindrical radius, $z$ is the cylindrical height, $r$ is the spherical radius, and
\begin{align}
g_R(R) &= 
\begin{cases}
    1,              & \qquad\text{if } R\leq \num{1.1}\rout, \\
    \exp \left(-\frac{(R-\num{1.1}\rout)^2}{2\cdot(\num{0.1}\rout)^2}\right)   & \qquad\text{otherwise},
\end{cases}\label{eqn:padding_gR}\\[10pt]
g_z(z) &= 
\begin{cases}
    1,              & \qquad\text{if } |z|\leq \num{1.1}H, \\
    \exp \left(-\frac{(|z|-\num{1.1}H)^2}{2\cdot(\num{0.1}H)^2}\right)   & \qquad\text{otherwise},
\end{cases}\\[10pt]
h(r) &= 
\begin{cases}
    \begin{aligned}[t]
    n_0 &+ 10(1-n_0)\left(\frac{x}{\rin}\right)^3 \\&- 15(1-n_0)\left(\frac{x}{\rin}\right)^4 + 6(1-n_0)\left(\frac{x}{\rin}\right)^5,    
    \end{aligned}
    & \parbox[t]{0.2\linewidth}{\vspace{18pt}$\text{if } r\leq \rin$,} \\[40pt]
    1   & \text{otherwise}.
\end{cases}
\end{align}
$p_0$ can be thought of as the normalization factor, and corresponds to the probability inside the volume $\rin \leq R<\rout, |z|<H$. The value is chosen so as to obtain the desired number of particles inside the shell. $g_R(R)g(z)$ provides the outer padding and begins dropping off as a Gaussian after a \SI{10}{\%} departure from the respective surface. The Gaussian shape of the roll-off is motivated by the normalizing flow having a Gaussian as the base distribution. Finally, $h(r)$ downsamples the virtual particles inside $r<\rin$, because the density of the stellar particles would be too large otherwise (tying back to the reason why the dataset was split into four shells in the first place). $h(r)$ drops from unity at $r=\rin$ down to $n_0$ at $r=0$, such that $h'(r=\rin)=h''(r=\rin)=h'(0)=h''(0)=0$. $n_0$ is chosen based on the density of stellar particles at the very center.

Finally, we create the datasets for each shell by selecting stellar particles with probability $p(\vec x)$, using the parameters listed in Table~\ref{table:shells}. In our 2 and \SI{4}{kpc} spherical sub-volumes (See Section~\ref{sec:solar_recovery}), we use $p\left(\vec{x}\right) = g_R\left(r\right)$, where $r$ is the spherical distance from the center of the given sub-volume, and $g_R$ has the form given above (Eq.~\ref{eqn:padding_gR}).



\bsp	
\label{lastpage}
\end{document}